\documentclass[a4paper,11pt]{article}
\usepackage{jheppub} 
\usepackage[utf8]{inputenc}
\usepackage{xspace}

\title{\boldmath Line-of-sight magnetic-field propagation effects on axion-like particle constraints from GRB 221009A}

\author{Chengcheng Han$^{1,2}$, }
\emailAdd{hanchch@mail.sysu.edu.cn}

\author{Zhanhong Lei$^1$, }
\emailAdd{leizhh3@mail2.sysu.edu.cn}

\author{Jiajie Yang$^1$, }
\emailAdd{yangjj68@mail2.sysu.edu.cn}

\author{and Shutong Zhao$^1$}
\emailAdd{zhaosht5@mail2.sysu.edu.cn}

\affiliation{$^1$School of Physics, Sun Yat-Sen University, Guangzhou 510275, P. R. China}
\affiliation{$^2$Asia Pacific Center for Theoretical Physics, Pohang 37673, Korea}

\abstract{High-energy photons from GRB 221009A provide a powerful opportunity to probe axion-like particles (ALPs) through photon-ALP oscillations in cosmic magnetic fields. We revisit the ALP constraints implied by the LHAASO observation of this burst, with particular emphasis on the magnetic-field environments encountered along the line of sight. We include the host-galaxy, intergalactic, and Milky-Way magnetic fields and assess their respective impacts on the photon survival probability and on the exclusion limits in the ALP mass-coupling plane. We show that the constraints are only mildly affected by the choice of host-galaxy and Galactic magnetic-field models, but can change significantly once the intergalactic magnetic field is varied. Its field strength, coherence scale, and stochastic properties can all leave visible imprints on the derived exclusion contours, and in some cases generate pronounced oscillatory features. This demonstrates that the intergalactic magnetic field constitutes the dominant astrophysical uncertainty in extracting ALP limits from GRB 221009A. Our analysis highlights the importance of realistic propagation modeling in future gamma-ray searches for ALPs.}

\begin{document}
\maketitle
\flushbottom

\section{Introduction}
\label{sec:intro}

Gamma-ray bursts (GRBs) are among the most luminous transient phenomena in the Universe, producing intense flashes of gamma rays with durations ranging from $\sim 10^{-2}$ to $\sim 10^{3}$ seconds. Owing to their enormous luminosities and cosmological distances, GRBs provide a unique laboratory for studying the propagation of high-energy photons over extragalactic scales. In particular, very-high-energy (VHE) photons from GRBs are expected to be strongly attenuated by the extragalactic background light (EBL) through the pair-production process $\gamma\gamma \rightarrow e^+ e^-$~\cite{gould1966opacity,Fazio:1970pr,protheroe2000infrared}. As a result, observations of VHE gamma rays from distant GRBs offer a sensitive probe of both photon propagation effects and possible new physics beyond the Standard Model.

GRB 221009A, the brightest long GRB ever recorded, was first detected by Fermi-GBM on 9 October 2022~\cite{lesage2023fermi}. Its source lies at redshift $z=0.151$, corresponding to a distance of about 753 Mpc~\cite{de2022grb}. Remarkably, LHAASO reported the detection of more than 5000 VHE photons from GRB 221009A, with energies extending up to 18 TeV~\cite{huang2022lhaaso,LHAASA:2023pay}. Such an observation is difficult to reconcile with conventional expectations based on standard EBL attenuation and is therefore in tension with most existing EBL models~\cite{dominguez2011extragalactic,saldana2021observational,finke2010modeling,finke2022modeling}. This tension may be related to systematic effects, such as the misidentification of cosmic rays as photons, uncertainties in energy reconstruction, or limitations of current EBL modeling~\cite{LHAASA:2023pay}. On the other hand, it may also indicate new physics mechanisms that enhance the transparency of the Universe to VHE photons. Among the possibilities discussed in the literature are Lorentz invariance violation~\cite{he2022lorentz} and axion-like particles (ALPs)~\cite{galanti2022axion}.

ALPs are light pseudoscalar bosons that appear in many extensions of the Standard Model. In a model-independent framework, they are mainly characterized by their mass $m_a$ and their two-photon coupling $g_{a\gamma\gamma}$~\cite{raffelt1988mixing,maiani1986effects,sikivie1983experimental,zhang2023axion}. In the presence of external magnetic fields, this coupling induces oscillations between photons and ALPs, thereby altering photon propagation in astrophysical environments. Since magnetic fields are ubiquitous on galactic and intergalactic scales, ALP-photon mixing can leave observable signatures in high-energy photon spectra. The same conversion mechanism has also been explored in related supernova-axion studies, including axion or ALP production from core-collapse supernovae, compact-star environments such as supernova remnants and neutron-star mergers, and the cosmic supernova population, followed by conversion into gamma rays in progenitor, host-galaxy, intergalactic, or Galactic magnetic fields~\cite{Fiorillo_2026,kanodia2026lightscameraaxiontracing,Cand_n_2026,fiorillo2026magneticturbulenceboostssupernova}. In particular, VHE photons emitted from a GRB may convert into ALPs in the magnetic field of the host galaxy, traverse intergalactic space with negligible EBL attenuation, and then partially reconvert into photons in the Milky Way magnetic field~\cite{de2007evidence}. This mechanism effectively increases the photon survival probability and can substantially modify the observed VHE spectrum~\cite{tavecchio2015photons,galanti2020hint,galanti2019blazar,de2011relevance,sanchez2009hints,simet2008milky}.

The TeV spectrum of GRB 221009A therefore provides an excellent opportunity to test the ALP interpretation and to constrain the ALP parameter space in the $m_a$--$g_{a\gamma\gamma}$ plane. For a fixed ALP mass, if the coupling is sufficiently large, the photon survival probability may be enhanced to a level that is incompatible with the observed spectrum, which in turn allows one to place an upper bound on $g_{a\gamma\gamma}$. However, the predicted ALP-photon oscillation effect depends sensitively on the magnetic fields encountered along the line of sight. Therefore, uncertainties in the host galaxy magnetic field (HGMF), the intergalactic magnetic field (IGMF), and the Milky Way magnetic field (MWMF) can directly affect the inferred constraints. Although a number of studies have used GRB 221009A to derive bounds on ALPs~\cite{LHAASA:2023pay,gao2024constraints,galanti2023observability}, the impact of different magnetic-field models has not yet been examined in a fully systematic way.

The main purpose of this work is to quantify how magnetic-field uncertainties affect the ALP constraints derived from GRB 221009A. To this end, we study ALP-photon propagation with different models for the HGMF, IGMF, and MWMF~\cite{baktash2022interpretation,LHAASA:2023pay,gonzalez2023grb,carenza2022alp,troitsky2024towards,levan2023first,kronberg1994extragalactic,grasso2001magnetic,galanti2018behavior,kartavtsev2017extragalactic,jansson2012new,adam2016planck,unger2024coherent}, and derive the corresponding upper limits in the $m_a$--$g_{a\gamma\gamma}$ plane from a minimum-$\chi^2$ fit to the LHAASO spectrum at the $95\%$ confidence level. We find that the HGMF and MWMF model dependence is relatively mild, while the dominant uncertainty arises from the IGMF. For sufficiently large $m_a$, ALP-photon conversion in intergalactic space becomes negligible, and the resulting constraints are nearly insensitive to the IGMF. By contrast, for small $m_a$, oscillations in the extragalactic medium can significantly enhance the photon survival probability and lead to substantially stronger bounds than those obtained from conventional EBL attenuation alone. In the intermediate-mass region, the exclusion contours develop characteristic oscillatory features. These results show that a realistic treatment of line-of-sight magnetic fields, especially the poorly known IGMF, is essential for deriving robust ALP constraints from GRB 221009A.

The remainder of this paper is organized as follows. In Sec.~\ref{sec:oscillation}, we briefly review ALP-photon oscillations and their impact on photon propagation in external magnetic fields. In Sec.~\ref{sec:propagation}, we present the magnetic-field models adopted for different astrophysical environments and discuss their effects on photon propagation. In Sec.~\ref{sec:constraint}, we derive and compare the resulting constraints on the ALP parameter space. Finally, we summarize our conclusions in Sec.~\ref{sec:conclude}.

\section{ALP-photon oscillation formalism}
\label{sec:oscillation}

In this section, we briefly review the ALP-photon mixing formalism relevant for high-energy photon propagation in external magnetic fields. We focus on the ingredients that are directly needed for our analysis of GRB 221009A, including the beam propagation equation, the oscillation behavior in a single magnetic domain, and the photon survival probability along the line of sight. More detailed discussions can be found in Refs.~\cite{galanti2023observability,galanti2018behavior,de2011relevance,davies2021relevance,meyer2014detecting}.

Axion-like particles (ALPs) are light pseudoscalar bosons that commonly arise as pseudo-Nambu--Goldstone bosons in extensions of the Standard Model. For the present purpose, only the ALP mass $m_a$ and the two-photon coupling $g_{a\gamma\gamma}$ are relevant. The corresponding effective Lagrangian is
\begin{equation}
\label{eq:2.1}
\mathcal{L}_{\mathrm{ALP}}
=
\frac{1}{2}\partial^{\mu}a\,\partial_{\mu}a
-\frac{1}{2}m_{a}^{2}a^{2}
-\frac{1}{4}g_{a\gamma\gamma}F_{\mu\nu}\tilde{F}^{\mu\nu}a
=
\frac{1}{2}\partial^{\mu}a\,\partial_{\mu}a
-\frac{1}{2}m_{a}^{2}a^{2}
+g_{a\gamma\gamma}\,\mathbf{E}\cdot\mathbf{B}\,a,
\end{equation}
where $a$ denotes the ALP field, $F^{\mu\nu}$ is the electromagnetic field-strength tensor, and $\tilde{F}^{\mu\nu}=\frac{1}{2}\epsilon^{\mu\nu\rho\sigma}F_{\rho\sigma}$ is its dual. Here $\mathbf{E}$ is the electric field of the propagating photon beam, while $\mathbf{B}$ denotes the external magnetic field. For a photon beam with wave vector $\mathbf{k}$, only the transverse magnetic-field component,
\begin{equation}
\mathbf{B}_T=\mathbf{B}-(\mathbf{B}\cdot\hat{\mathbf{k}})\hat{\mathbf{k}},
\end{equation}
enters the photon-ALP mixing.

\subsection{Beam propagation equation}

Consider a photon-ALP beam of energy $E$ propagating along the $y$-direction in an external magnetic field. In the short-wavelength approximation, the beam evolution can be described by a Schr\"odinger-like equation~\cite{raffelt1988mixing},
\begin{equation}
\label{eq:schrodinger_new}
\left(i\frac{d}{dy}+E+\mathcal{M}(E,y)\right)\psi(y)=0,
\end{equation}
where $\mathcal{M}(E,y)$ is the mixing matrix, and
\begin{equation}
\psi(y)=\bigl(A_x(y),\,A_z(y),\,a(y)\bigr)^T
\end{equation}
is the photon-ALP state vector. Here $A_x(y)$ and $A_z(y)$ denote the photon polarization amplitudes along the $x$- and $z$-axes, respectively, and $a(y)$ is the ALP amplitude. Denoting by $U(E;y,y_0)$ the transfer matrix associated with Eq.~\eqref{eq:schrodinger_new}, one has
\begin{equation}
\label{eq:2.5}
\psi(y)=U(E;y,y_0)\,\psi(y_0),
\end{equation}
with the initial condition
\begin{equation}
U\bigl(E; y_{0}, y_{0}) = \mathbb{I}.
\end{equation}
Furthermore, one can set
\begin{equation}
\begin{aligned}
\label{eq:2.7}
U\bigl(E; y, y_{0})
= e^{iE (y-y_{0})}\,
\mathcal{U} \bigl(E; y, y_{0})
\end{aligned}
\end{equation}
where $\mathcal{U} \bigl(E; y, y_{0})$ is the transfer matrix associated with the reduced Schr\"odinger-like equation, 
\begin{equation}
\begin{aligned}
\label{eq:2.8}
\left( i \frac{d}{dy} + \mathcal{M}(E, y) \right)\psi(y)=0
\end{aligned}
\end{equation}
with the initial condition
\begin{equation}
\mathcal{U}\bigl(E; y_{0}, y_{0}) = \mathbb{I}.
\end{equation}
The mixing matrix can be written as~\cite{galanti2023observability}
\begin{equation}
\label{eq:2.10}
\mathcal{M}(E,y)
=
V^\dagger(\phi)
\begin{pmatrix}
\Delta_{\perp}(E,y) & 0 & 0 \\
0 & \Delta_{\parallel}(E,y) & \Delta_{a\gamma}(y) \\
0 & \Delta_{a\gamma}(y) & \Delta_{aa}(E)
\end{pmatrix}
V(\phi),
\end{equation}
where $\phi$ is the angle between $\mathbf{B}_T$ and the $z$-axis, and $V(\phi)$ is the rotation matrix in the $x$-$z$ plane.
\begin{equation}
\begin{aligned}
V(\phi)=
\begin{pmatrix}
\cos\phi & -\sin\phi & 0\\
\sin\phi & \cos\phi & 0\\
0 & 0 & 1
\end{pmatrix}
\end{aligned}
\end{equation}
The diagonal photon terms are
\begin{equation}
\begin{aligned}
\Delta_{\perp}(E,y)
&=
\Delta_{\mathrm{pl}}(E,y)
+2\Delta_{\mathrm{QED}}(E,y)
+\Delta_{\mathrm{CMB}}(E)
+\frac{i}{2\lambda_{\gamma}(E,y)},
\\
\Delta_{\parallel}(E,y)
&=
\Delta_{\mathrm{pl}}(E,y)
+\frac{7}{2}\Delta_{\mathrm{QED}}(E,y)
+\Delta_{\mathrm{CMB}}(E)
+\frac{i}{2\lambda_{\gamma}(E,y)}.
\end{aligned}
\end{equation}
The various entries in Eq.~\eqref{eq:2.10} have the following physical meaning:
\begin{itemize}
\item
\begin{equation}
\Delta_{a\gamma}(y)=\frac{1}{2}g_{a\gamma\gamma}B_T(y),
\end{equation}
which describes photon-ALP mixing induced by the external magnetic field.

\item
\begin{equation}
\Delta_{aa}(E)=-\frac{m_a^2}{2E},
\end{equation}
which encodes the ALP mass contribution.

\item
\begin{equation}
\Delta_{\mathrm{pl}}(E,y)=-\frac{\omega_{\mathrm{pl}}^2(y)}{2E},
\end{equation}
which is the plasma contribution arising from the effective photon mass in a cold medium. Here the plasma frequency is
\begin{equation}
\omega_{\mathrm{pl}}(y)=\left(\frac{4\pi\alpha n_e(y)}{m_e}\right)^{1/2},
\end{equation}
with $n_e$ and $m_e$ being the electron number density and electron mass, respectively.

\item
\begin{equation}
\Delta_{\mathrm{QED}}(E,y)=\frac{\alpha E}{45\pi}\left(\frac{B_T(y)}{B_{\mathrm{cr}}}\right)^2,
\end{equation}
which accounts for the QED vacuum polarization effect~\cite{heisenberg1936folgerungen,weisskopf1936elektrodynamik,schwinger1951gauge}. Here $B_{\mathrm{cr}}\simeq 4.41\times10^{13}\,\mathrm{G}$ is the critical magnetic field.

\item
\begin{equation}
\Delta_{\mathrm{CMB}}(E)=\chi_{\mathrm{CMB}}E,
\end{equation}
which describes photon dispersion on the cosmic microwave background, with $\chi_{\mathrm{CMB}}=0.522\times10^{-42}$~\cite{dobrynina2015photon}.

\item
\begin{equation}
\frac{i}{2\lambda_\gamma(E,y)},
\end{equation}
which accounts for photon absorption due to the process $\gamma\gamma\to e^+e^-$ on background radiation fields. Here $\lambda_\gamma(E,y)$ is the photon mean free path.
\end{itemize}

Among these terms, $\Delta_{a\gamma}$, $\Delta_{aa}$, and $\Delta_{\mathrm{pl}}$ usually determine the basic mixing structure, while $\Delta_{\mathrm{QED}}$ becomes important in sufficiently strong magnetic fields and $\Delta_{\mathrm{CMB}}$ can play a significant role in the high-energy behavior during intergalactic propagation. Their interplay controls the energy dependence of the photon survival probability.

Since the polarization state of the emitted VHE photons is generally unknown and the final photon polarization is not measured, it is convenient to describe the beam by a density matrix rather than by a pure-state wave function. We therefore introduce
\begin{equation}
\rho(y)=\psi(y)\otimes\psi^\dagger(y),
\end{equation}
whose evolution is governed by the von Neumann-like equation
\begin{equation}
\label{eq:von_neumann_new}
i\frac{d\rho(y)}{dy}
=
\rho(y)\mathcal{M}^{\dagger}(E,y)
-
\mathcal{M}(E,y)\rho(y).
\end{equation}
The formal solution is
\begin{equation}
\label{eq:2.21}
\rho(y)=\mathcal{U}(E;y,y_0)\,\rho_0\,\mathcal{U}^{\dagger}(E;y,y_0),
\end{equation}
where $\rho_0$ is the initial density matrix. The probability for a beam prepared in the state $\rho_1$ at $y_0$ to be observed in the state $\rho_2$ at $y$ is then
\begin{equation}
P_{\rho_1\to\rho_2}(E,y)
=
\mathrm{Tr}
\left[
\rho_2\,
\mathcal{U}(E;y,y_0)\,
\rho_1\,
\mathcal{U}^{\dagger}(E;y,y_0)
\right],
\end{equation}
with $\mathrm{Tr}\,\rho_1=\mathrm{Tr}\,\rho_2=1$.

\subsection{Oscillation length and conversion probability in a single magnetic domain}

The full beam evolution in a realistic astrophysical environment is generally complicated. Nevertheless, useful analytical insight can be obtained in an idealized single-domain setup. To this end, we temporarily neglect photon absorption and consider propagation through a magnetic domain of size $L_{\mathrm{dom}}$ with homogeneous medium and uniform transverse magnetic field. 
In this setup, only the photon polarization state whose polarization direction is parallel to the external transverse magnetic field $B_T$ mixes with the ALP, while the orthogonal photon polarization state does not participate in photon--ALP oscillations. The conversion probability for the parallel polarization state is given by~\cite{galanti2018behavior},
\begin{equation}
\label{eq:2.24}
P_{\gamma\to a}(E,y)
=
\left(
\frac{g_{a\gamma\gamma}B_T\,l_{\mathrm{osc}}(E)}{2\pi}
\right)^2
\sin^2\left(\frac{\pi y}{l_{\mathrm{osc}}(E)}\right),
\qquad y\leq L_{\mathrm{dom}},
\end{equation}
where the oscillation length is
\begin{equation}
\label{eq:losc}
l_{\mathrm{osc}}(E)
=
2\pi
\left[
\left\{
\frac{m_a^2-\omega_{\mathrm{pl}}^2}{2E}
+
\left(
1.42\times10^{-4}\left(\frac{B_T}{B_{\mathrm{cr}}}\right)^2
+
0.522\times10^{-42}
\right)E
\right\}^2
+
\left(g_{a\gamma\gamma}B_T\right)^2
\right]^{-1/2}.
\end{equation}

To clarify the energy dependence of $l_{\mathrm{osc}}(E)$, we identify the three contributions entering Eq.~\eqref{eq:losc}. The first term inside the braces is the mass-dependent contribution, including the ALP mass and the plasma frequency, and scales as $E^{-1}$. The second term is proportional to $E$ and contains two contributions: the QED vacuum-polarization contribution, proportional to $(B_T/B_{\mathrm{cr}})^2$, and the CMB dispersion contribution, proportional to $0.522\times10^{-42}$. The last term, $(g_{a\gamma\gamma}B_T)^2$, is the energy-independent photon--ALP mixing contribution induced by the external transverse magnetic field. Since these three contributions have different energy dependences, the explicit form of $l_{\rm osc}(E)$ simplifies whenever one of them dominates over the others.

Through Eq.~\eqref{eq:2.24}, the energy dependence of $l_{\rm osc}(E)$ also controls the behavior of the single-domain conversion probability $P_{\gamma\to a}(E,L_{\rm dom})$. We therefore discuss below the approximate forms of $l_{\rm osc}(E)$ in different energy regimes, from which the corresponding behavior of $P_{\gamma\to a}$ can be inferred. For this purpose, it is useful to define the characteristic low- and high-energy scales~\cite{galanti2018behavior},
\begin{equation}
\begin{aligned}
E_L
&\equiv
\frac{|m_a^2-\omega_{\mathrm{pl}}^2|}{2g_{a\gamma\gamma}B_T},
\\
E_H
&\equiv
g_{a\gamma\gamma}B_T
\left[
1.42\times10^{-4}\left(\frac{B_T}{B_{\mathrm{cr}}}\right)^2
+
0.522\times10^{-42}
\right]^{-1}.
\end{aligned}
\end{equation}
If $E_L \lesssim E_H$, these scales separate three qualitatively different regimes:

\begin{itemize}
\item
\textbf{Low-energy weak-mixing regime:} $E \lesssim E_L$. In this regime, the term proportional to $E^{-1}$ dominates in Eq.~\eqref{eq:losc}. Thus, we have
\begin{equation}
\label{eq:low-energy}
l_{\rm osc}(E)
\simeq 
\frac{4\pi E}{|m_a^2-\omega_{\rm pl}^2|}.
\end{equation}
The oscillation length $l_{\rm osc}$ is energy dependent and increases with energy. $P_{\gamma\to a}(E,L_{\rm dom})$ also increases with energy, while the $\sin^2(\pi L_{\rm dom}/l_{\rm osc})$ factor can induce oscillatory behavior in $E$.

\item
\textbf{Strong-mixing regime:} $E_L \lesssim E\lesssim E_H$. In this regime, the energy-independent term $(g_{a\gamma\gamma}B_T)^2$ dominates in Eq.~\eqref{eq:losc}. Accordingly, we have
\begin{equation}
\label{eq:strong-mixing}
l_{\rm osc}
\simeq
\frac{2\pi}{g_{a\gamma\gamma}B_T},
\end{equation}
and
\begin{equation}
P_{\gamma\to a}(L_{\rm dom})
\simeq
\sin^2\left(
\frac{g_{a\gamma\gamma}B_TL_{\rm dom}}{2}
\right).
\end{equation}
In this case, both $l_{\rm osc}$ and $P_{\gamma\to a}$ are approximately independent of $m_a$ and $E$, and the conversion probability reaches its maximal amplitude.

\item
\textbf{High-energy weak-mixing regime:} $E>E_H$. 
This is the regime where the term proportional to $E$ dominates in Eq.~\eqref{eq:losc}. Accordingly, we obtain
\begin{equation}
\label{eq:high-energy}
l_{\rm osc}(E) \simeq \frac{2\pi}{E} 
\left[1.42\times10^{-4}\left(\frac{B_T}{B_{\rm cr}}\right)^2
+0.522\times10^{-42}\right]^{-1}.
\end{equation}
$l_{\rm osc}(E)$ and $P_{\gamma\to a}(E,L_{\rm dom})$ decrease as $E$ increases, while the $\sin^2$ factor can again induce oscillatory behavior in $P_{\gamma\to a}(E,L_{\rm dom})$ with energy.
\end{itemize}

Another possibility is $E_L \gtrsim E_H$. In this case, the sum of the term proportional to $E^{-1}$ and the term proportional to $E$ inside the braces in Eq.~\eqref{eq:losc} is larger than $g_{a\gamma\gamma}B_T$ over the whole energy range. Therefore, the $(g_{a\gamma\gamma}B_T)^2$ term does not dominate in Eq.~\eqref{eq:losc}. As a result, the strong-mixing regime is absent, and the low-energy and high-energy weak-mixing regimes are connected directly. Consequently, $l_{\rm osc}(E)$ first increases with energy, reaches a maximum, and then decreases at higher energies.

\begin{figure}[htbp]
\centering
\includegraphics[width=0.75\textwidth]{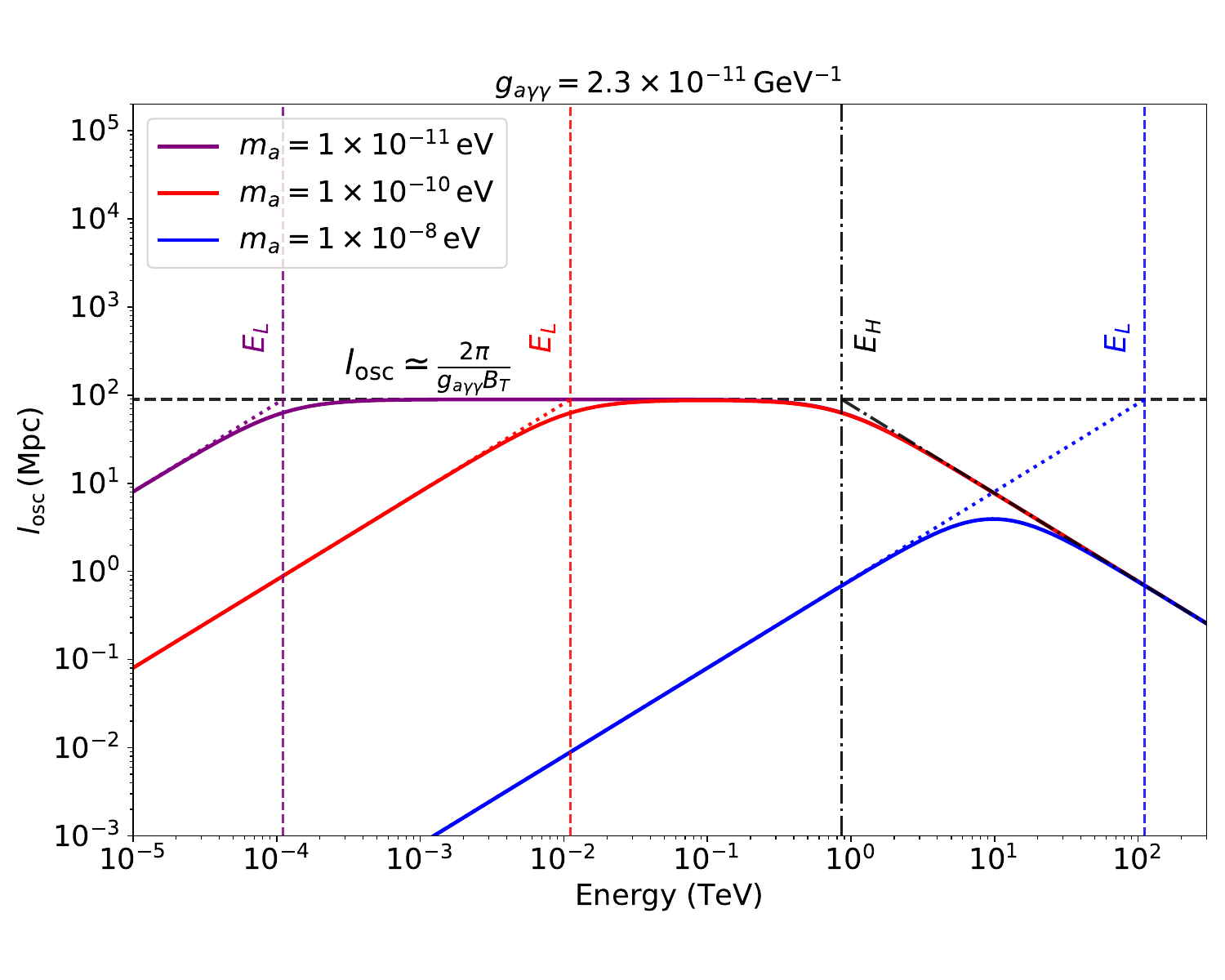}
\caption{Energy dependence of the single-domain oscillation length $l_{\rm osc}(E)$ for three representative ALP masses, $m_a=1\times10^{-11}\,\mathrm{eV}$, $1\times10^{-10}\,\mathrm{eV}$, and $1\times10^{-8}\,\mathrm{eV}$. The benchmark parameters are $g_{a\gamma\gamma}=2.3\times10^{-11}\,\mathrm{GeV}^{-1}$, $B_T=1\,\mathrm{nG}$, and $n_e\sim10^{-7}\,\mathrm{cm}^{-3}$. The vertical dashed lines mark $E_H$ and the corresponding $E_L$ for each ALP mass, while the horizontal, black oblique, and colored oblique dashed lines indicate the strong-mixing, high-energy weak-mixing, and low-energy weak-mixing approximations, respectively.}
\label{fig:1}
\end{figure}

For illustration, we use the IGMF benchmark parameters to show the single-domain behavior of $l_{\rm osc}(E)$. Figure~\ref{fig:1} shows the energy dependence of the oscillation length for three representative ALP masses, $m_a=1\times10^{-11}\,\mathrm{eV}$, $1\times10^{-10}\,\mathrm{eV}$, and $1\times10^{-8}\,\mathrm{eV}$, corresponding to the purple, red, and blue solid curves, respectively. We use $B_T=1\,\mathrm{nG}$, $g_{a\gamma\gamma}=2.3\times10^{-11}\,\mathrm{GeV}^{-1}$, and $n_e\sim10^{-7}\,\mathrm{cm}^{-3}$. In the parameter range shown in Fig.~\ref{fig:1}, one has $m_a^2\gg\omega_{\rm pl}^2$, so the plasma contribution to the mass-dependent term is negligible.

The vertical black dashed line marks the universal high-energy scale $E_H$, whereas the colored vertical dashed lines mark the corresponding low-energy scales $E_L$ for each ALP mass. The horizontal black dashed line indicates the value of $l_{\rm osc}$ in the energy-independent strong-mixing regime, where $l_{\rm osc}\simeq 2\pi/(g_{a\gamma\gamma}B_T)$. The figure shows how the relative ordering of $E_L$ and $E_H$ controls the energy dependence of $l_{\rm osc}$. For the two smaller masses considered here, $m_a=1\times10^{-11}\,\mathrm{eV}$ and $1\times10^{-10}\,\mathrm{eV}$, the ordering is $E_L<E_H$. In this case, when $E<E_L$, $l_{\rm osc}$ keeps increasing with energy until it enters the strong-mixing regime, where $l_{\rm osc}$ is approximately constant. When $E>E_H$, $l_{\rm osc}$ starts to decrease with energy. For $m_a=1\times10^{-8}\,\mathrm{eV}$, one has $E_L>E_H$, and the energy-independent strong-mixing regime is absent. The low-energy and the high-energy weak-mixing regime are then connected directly, so $l_{\rm osc}$ first increases with energy, reaches a maximum, and then decreases at higher energies. The oblique black dashed line corresponds to the high-energy weak-mixing approximation in Eq.~\eqref{eq:high-energy}. The colored oblique dashed lines represent the low-energy weak-mixing approximation in Eq.~\eqref{eq:low-energy} for the corresponding ALP masses. They indicate the approximate behaviors discussed above.

\begin{figure}[htbp]
\centering
\includegraphics[width=0.75\textwidth]{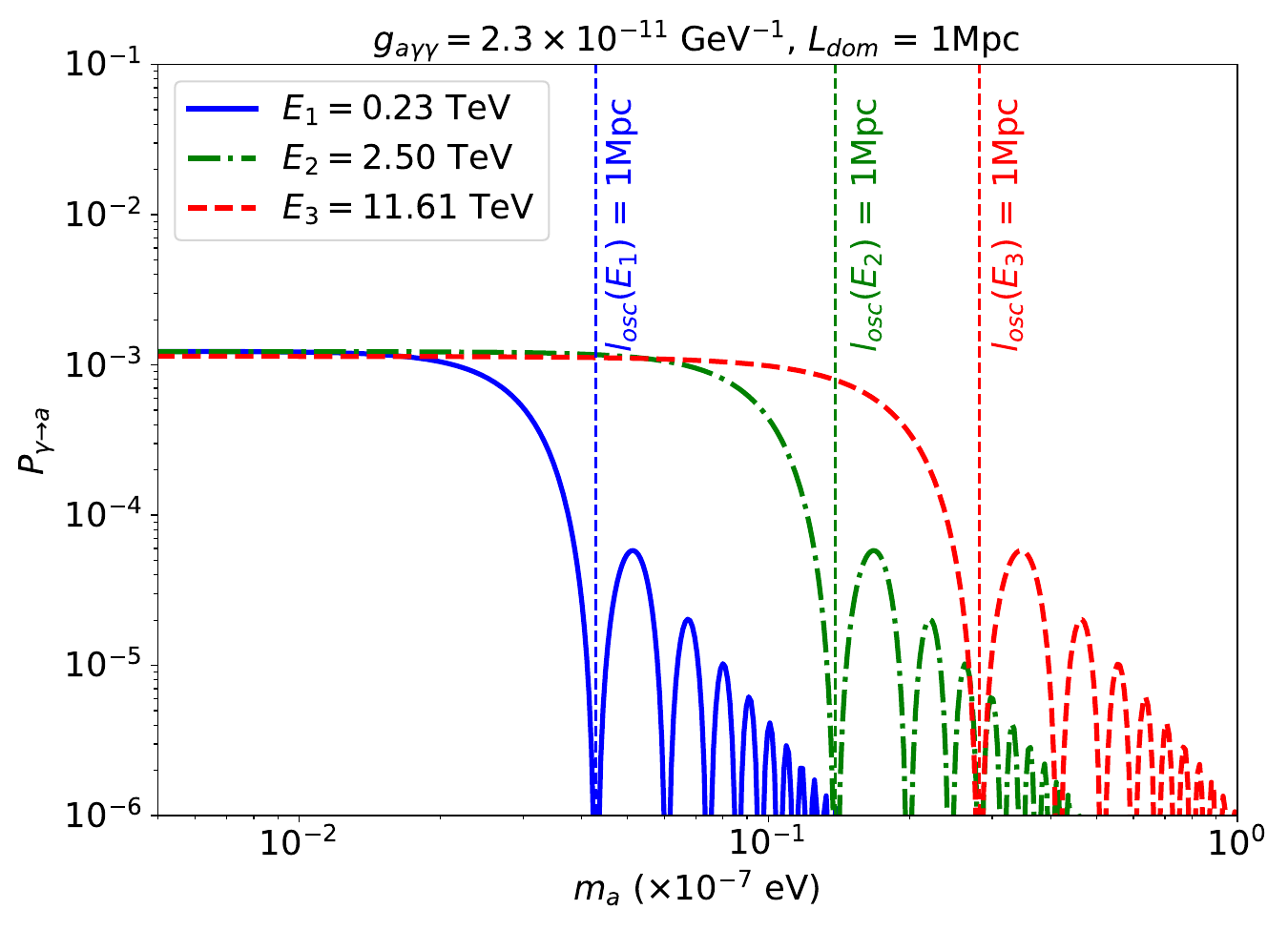}
\caption{Photon-ALP conversion probability $P_{\gamma\to a}$ in a single magnetic domain, plotted against the ALP mass $m_a$. The probability is shown for the photon polarization state parallel to the external transverse magnetic field $B_T$. The benchmark parameters are $L_{\rm dom}=1\,\mathrm{Mpc}$, $B_T=1\,\mathrm{nG}$, $g_{a\gamma\gamma}=2.3\times10^{-11}\,\mathrm{GeV}^{-1}$, and $n_e\sim10^{-7}\,\mathrm{cm}^{-3}$. The three curves correspond to $E=0.23$, $2.50$, and $11.61\,\mathrm{TeV}$. The vertical dashed lines mark the masses at which $l_{\rm osc}(E)=L_{\rm dom}$ for the corresponding photon energies.}

\label{fig:2}
\end{figure}

We next show the single-domain $\gamma\to a$ conversion probability. Figure~\ref{fig:2} shows the $\gamma\to a$ conversion probability $P_{\gamma\to a}$ in a single magnetic domain for the photon polarization state whose polarization direction is parallel to the external transverse magnetic field $B_T$, plotted against the ALP mass $m_a$. We use the same IGMF benchmark parameters as in Fig.~\ref{fig:1}, namely $B_T=1\,\mathrm{nG}$, $g_{a\gamma\gamma}=2.3\times10^{-11}\,\mathrm{GeV}^{-1}$, and $n_e\sim10^{-7}\,\mathrm{cm}^{-3}$, and set the domain length to $L_{\rm dom}=1\,\mathrm{Mpc}$. The three curves correspond to $E=0.23$, $2.50$, and $11.61\,\mathrm{TeV}$, respectively, which cover the energy range used in the following analysis. The vertical dashed lines mark the masses at which $l_{\rm osc}(E)=L_{\rm dom}$ for the corresponding photon energies.

The figure shows that the single-domain conversion probability is strongly suppressed at large $m_a$. In this region, the oscillation length is short and the prefactor in Eq.~\eqref{eq:2.24}, $(g_{a\gamma\gamma}B_Tl_{\rm osc}/2\pi)^2$, is small, while the sine factor induces rapid oscillations in the conversion probability. As $m_a$ decreases and $l_{\rm osc}$ becomes comparable to $L_{\rm dom}$, the envelope of the conversion probability increases rapidly. For even smaller masses, $l_{\rm osc}$ further increases and eventually becomes much larger than $L_{\rm dom}$. In this limit, the conversion probability approaches an approximately mass-independent and energy-independent constant.

\subsection{Photon survival probability through multiple magnetized environments}

For photons emitted by GRB 221009A, the beam propagates through multiple magnetized environments before reaching the Earth, including the host galaxy, the intergalactic medium, and the Milky Way. The full propagation path can therefore be divided into a sequence of magnetic domains, and the total transfer matrix is given by the ordered product
\begin{equation}
\mathcal{U}(E;y_{N+1},y_1)
=
\prod_{i=1}^{N}
\mathcal{U}_i(E;y_{i+1},y_i),
\end{equation}
where $\mathcal{U}_i(E;y_{i+1},y_i)$ denotes the transfer matrix across the $i$-th domain, and $y_1$ and $y_{N+1}$ are the source and detector positions, respectively.

Assuming that the emitted photons are initially unpolarized, we take
\begin{equation}
\rho_0=\mathrm{diag}\left(\frac{1}{2},\,\frac{1}{2},\,0\right).
\end{equation}
The total photon survival probability is then
\begin{equation}
\label{eq:2.28}
P_{\mathrm{ALP}}(E;\gamma\to\gamma)
=
\sum_{i=x,z}
\mathrm{Tr}
\left[
\rho_i\,
\mathcal{U}(E;y_{N+1},y_1)\,
\rho_0\,
\mathcal{U}^{\dagger}(E;y_{N+1},y_1)
\right],
\end{equation}
where the sum runs over the two final photon polarization states,
\begin{equation}
\rho_x=\mathrm{diag}(1,0,0),
\qquad
\rho_z=\mathrm{diag}(0,1,0).
\end{equation}

Equation~\eqref{eq:2.28} shows that the photon survival probability is controlled by the total transfer matrix along the line of sight. Besides the ALP parameters $(m_a,g_{a\gamma\gamma})$, this transfer matrix depends on the astrophysical inputs that determine photon--ALP propagation in each region, including the magnetic-field strength, orientation, and characteristic length scale, the electron density, and the photon mean free path. Therefore, different assumptions about the host-galaxy, intergalactic, and Galactic magnetic fields can lead to appreciable differences in the inferred ALP constraints. This is the main motivation for the systematic comparison of magnetic-field models in the following section.

\section{Propagation of photons from GRB 221009A through different astrophysical regions}
\label{sec:propagation}

Photons emitted by GRB 221009A propagate through three main astrophysical environments before reaching the Earth: the host galaxy (HG), intergalactic space, and the Milky Way (MW). In the presence of ALPs, photons may convert into ALPs in the host-galaxy magnetic field, travel through extragalactic space with reduced attenuation, and then partially reconvert into photons in the Galactic magnetic field. The corresponding beam evolution depends on the properties of the medium and magnetic field in each region. We therefore specify below the astrophysical setups and magnetic-field models adopted in our numerical analysis.

\subsection{Host galaxy}

Hubble Space Telescope observations indicate that the host galaxy of GRB 221009A is a disk-like galaxy viewed nearly edge-on, with the burst located close to its central region~\cite{levan2023first,troitsky2024towards}. Since coherent magnetic fields with typical strengths of order $\mu{\rm G}$ are commonly observed in disk galaxies on kpc scales~\cite{govoni2004magnetic,feretti2012clusters}, photon-ALP conversion inside the host galaxy can be relevant for the propagation of VHE photons. For all host-galaxy magnetic-field models considered below, we describe the electron density using the NE2001 model~\cite{cordes2002ne2001}, following Ref.~\cite{gao2024constraints}.

We consider three representative models for the host-galaxy magnetic field (HGMF):
\begin{itemize}
\item \textbf{Model 1:} a simplified homogeneous-field model adopted in Refs.~\cite{baktash2022interpretation,LHAASA:2023pay}, with transverse magnetic field $B_T\simeq0.5\,\mu{\rm G}$ and coherence length $\sim10\,{\rm kpc}$.

\item \textbf{Model 2:} a phenomenological proxy in which the host-galaxy magnetic-field configuration is assumed to be identical to that of the Milky Way~\cite{gonzalez2023grb,carenza2022alp};

\item \textbf{Model 3:} a more realistic model constructed in Ref.~\cite{troitsky2024towards}, based on HST observations of the host galaxy~\cite{levan2023first} together with magnetic-field measurements and simulations for NGC 891-like galaxies. 
\end{itemize}

\begin{figure}[htbp]
\centering
\includegraphics[width=0.47\textwidth]{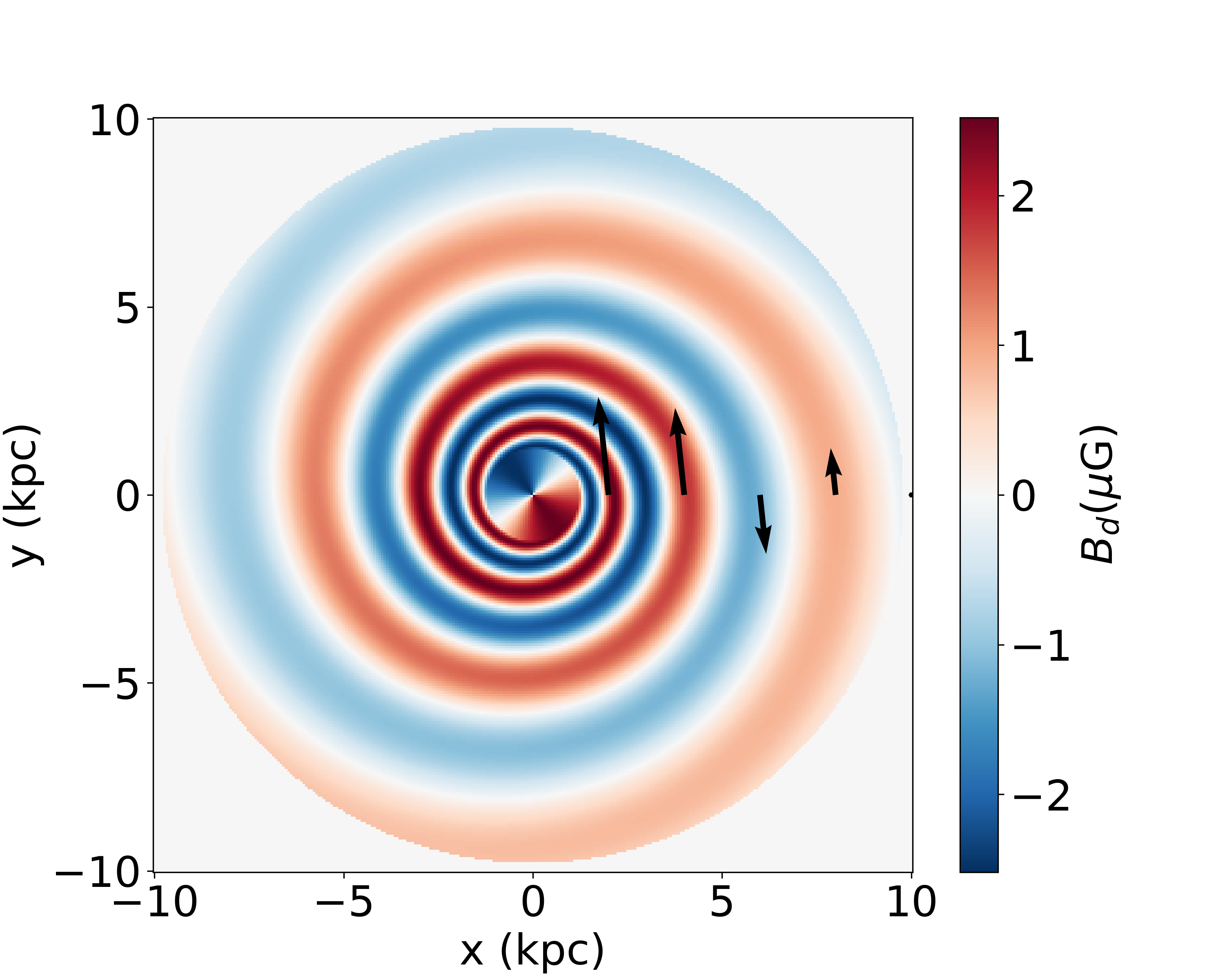}
\qquad
\includegraphics[width=0.47\textwidth]{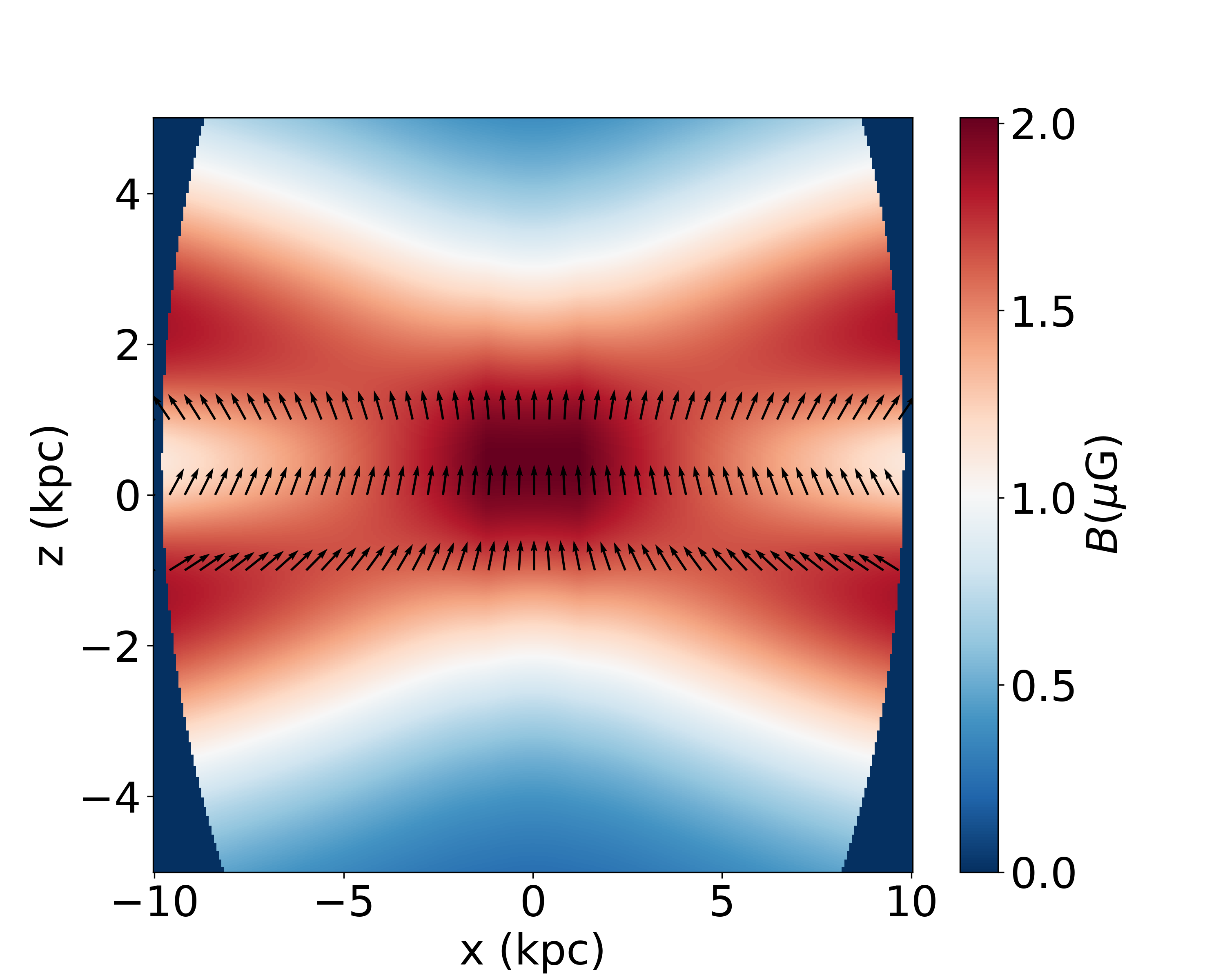}
\caption{Disk magnetic field (left) and X-shaped halo magnetic field (right) in the host-galaxy Model~3. In the left panel, the color map distinguishes the two opposite orientations of the disk field along the spiral pattern: red and blue regions correspond to opposite field directions, while the color intensity indicates the local field amplitude. In the right panel, the color map represents the magnetic-field strength of the X-shaped halo component. In both panels, the black arrows indicate the projected local direction of the magnetic field.}
\label{fig:A1}
\end{figure}

Figure~\ref{fig:A1} illustrates the magnetic-field structure of the host-galaxy Model~3. Here we use a host-galaxy-centered Cartesian coordinate system, in which the $x$-$y$ plane coincides with the galactic disk and the $z$-axis is perpendicular to the disk. This coordinate system is used only for specifying the spatial morphology of the host-galaxy magnetic field; in particular, its $y$-axis is not the same as the propagation coordinate $y$ introduced in Sec.~\ref{sec:oscillation} for the photon--ALP beam. The left panel shows the disk component in the host galaxy, where the color map distinguishes the two opposite field orientations along the spiral pattern, with red and blue regions corresponding to opposite field directions and the color intensity indicating the local field amplitude. The right panel shows the X-shaped halo component in a vertical cross section, with the color map representing the magnetic-field strength. In both panels, the black arrows indicate the local magnetic-field direction projected onto the corresponding plane.

For Model 3, the propagation also depends on the burst position inside the host galaxy. Following Refs.~\cite{troitsky2024towards,gao2024constraints}, we characterize this dependence by two geometric parameters: the line-of-sight position $y_0$ of GRB 221009A and the polar angle $\theta_0$ in the galactic disk plane. Since the resulting ALP constraints depend only weakly on these parameters~\cite{gao2024constraints}, we fix them to the benchmark values
\begin{equation}
(\theta_0,y_0)=(0^\circ,\,0~{\rm kpc}).
\end{equation}

\subsection{Extragalactic space}

After leaving the host galaxy, VHE photons propagate through extragalactic space and are attenuated by pair production on the extragalactic background light (EBL), $\gamma\gamma\to e^+e^-$. In order to remain consistent with the LHAASO analysis, we adopt the EBL model of Saldana-Lopez et al.~\cite{saldana2021observational} and compute the corresponding photon mean free path $\lambda_\gamma(E,y)$. We have checked that alternative EBL models lead to only minor differences within the energy range relevant for this work~\cite{Meyer2022}.

The role of the intergalactic magnetic field (IGMF) is more uncertain. Because the IGMF is weak and poorly constrained, ALP-photon oscillations in extragalactic space are often neglected. In the present work, however, we include this effect explicitly, since it can have an important impact on the photon survival probability and hence on the inferred ALP constraints. Current observations suggest
\begin{equation}
10^{-7}\,{\rm nG}\lesssim B_{\rm IG}\lesssim1.7\,{\rm nG}
\end{equation}
on Mpc scales~\cite{neronov2010evidence,durrer2013cosmological,pshirkov2016new}, 
while recent studies indicate that a substantial fraction of the intergalactic volume may be occupied by magnetized filamentary regions, with filling fraction $f\gtrsim0.67$ in a simplified top-hat model~\cite{tjemsland2024constraining}. 

Motivated by these considerations, we model the IGMF as a sequence of randomly oriented magnetized domains interspersed with field-free voids. Each magnetized domain has size $L_{\rm dom}$, identified with the IGMF coherence length. If $D$ denotes the distance between the centers of two adjacent magnetized domains, then the void size is $D-L_{\rm dom}$ and the filling fraction is
\begin{equation}
f=\frac{L_{\rm dom}}{D}.
\end{equation}
The IGMF is thus characterized by the parameters $B_{\rm IG}$, $L_{\rm dom}$, and $f$, and photon-ALP propagation in extragalactic space becomes a stochastic process. To assess how large an effect this process can produce and to facilitate an analytic treatment of the underlying mechanism, we adopt a typical benchmark with $B_{\rm IG}\sim\mathcal{O}(1)\,\mathrm{nG}$ and use the averaged transverse component $B_T=\sqrt{2/3}\,B=1\,\mathrm{nG}$. This corresponds to averaging over randomly oriented magnetic domains, since only the transverse magnetic field couples to ALPs. The same effective value of $B_T$ is then assigned to each individual domain. For the intergalactic medium, we fix the electron density to $n_e\sim10^{-7}\,\mathrm{cm}^{-3}$~\cite{de2011relevance}.

Within this framework, we consider two standard realizations:
\begin{itemize}
\item \textbf{DLSHE model:} in the domain-like sharp-edges model (DLSHE)~\cite{kronberg1994extragalactic,grasso2001magnetic}, the magnetic field is homogeneous inside each domain, while its orientation changes randomly and discontinuously at domain boundaries. This approximation is computationally simple, but may become unreliable once photon dispersion on the CMB is important~\cite{dobrynina2015photon}. Its validity requires
\begin{equation}
l_{\rm osc}(E)\gg L_{\rm dom},
\end{equation}
so that the conversion probability is insensitive to the detailed domain profile.

\item \textbf{DLSME model:} in the domain-like smooth-edges model (DLSME)~\cite{galanti2018behavior,kartavtsev2017extragalactic}, the magnetic-field components vary continuously across neighboring domains. The smoothness of the directional transition is parameterized by $\sigma\in[0,1]$, with the limiting case $\sigma=0$ reproducing the DLSHE setup.
\end{itemize}

As shown in Sec.~\ref{sec:constraint}, the IGMF introduces the largest astrophysical uncertainty in the ALP constraints derived from GRB 221009A.

\subsection{Milky Way}

Before reaching the Earth, the photon-ALP beam traverses the Milky Way magnetic field (MWMF), where ALPs may convert back into photons. The Milky Way magnetic field is commonly decomposed into a large-scale regular component and a small-scale turbulent component. The latter typically has coherence length $\lambda\lesssim100~{\rm pc}$~\cite{haverkorn2008outer,gaensler1995pulsar}, which is too short to induce sizable ALP-photon oscillations at the gamma-ray energies of interest. We therefore retain only the regular component. We note, however, that the turbulent component may be relevant in other ALP--photon conversion searches, in particular for MeV gamma-ray signals from supernova axions, where it can modify the conversion probability and extend the resulting sensitivity to higher axion masses~\cite{fiorillo2026magneticturbulenceboostssupernova}.

For the electron density distribution, we adopt the NE2001 model~\cite{cordes2002ne2001}. Following Ref.~\cite{gao2024constraints}, we also use it as an effective description of the electron density in the host galaxy. For the regular Galactic magnetic field, we consider two classes of models:
\begin{itemize}
\item \textbf{JF models:} the Jansson--Farrar model JF12~\cite{jansson2012new} and its updated versions JF12b and JF12c~\cite{adam2016planck};

\item \textbf{UF models:} the more recent Unger--Farrar models~\cite{unger2024coherent}, whose parameters are fitted to updated full-sky Faraday rotation measures of extragalactic sources and polarized synchrotron maps from WMAP and Planck. We consider the eight benchmark realizations \texttt{base}, \texttt{expX}, \texttt{spur}, \texttt{neCL}, \texttt{twistX}, \texttt{nebCor}, \texttt{cre10}, and \texttt{synCG}.
\end{itemize}

In the following, we compare the ALP constraints obtained with these different HGMF, IGMF, and MWMF models, in order to assess which magnetic-field uncertainties are most relevant for photon-ALP propagation from GRB 221009A.

\section{Constraints on ALP parameters}
\label{sec:constraint}
Based on the propagation framework described above, we derive constraints on the ALP parameter space using the observed VHE spectrum of GRB 221009A from LHAASO~\cite{LHAASA:2023pay}. GRB 221009A was initially detected by Fermi-GBM at $T_0$. The LHAASO data are divided into two time intervals according to the light curve observed by KM2A. The first interval spans from $T_0+230~{\rm s}$ to $T_0+300~{\rm s}$, while the second interval extends from $T_0+300~{\rm s}$ to $T_0+900~{\rm s}$. As a simplified benchmark analysis, we use the spectral data from the first time interval, $T_0+230~{\rm s}$ to $T_0+300~{\rm s}$, whose energy range is $0.23\,\mathrm{TeV}$ to $11.61\,\mathrm{TeV}$, to derive the ALP constraints. 
The numerical computation of ALP-photon oscillations in the host galaxy, intergalactic space, and the Milky Way is performed with the open-source code $\mathtt{gammaALPs}$~\cite{meyer2021gammaalps,meyer2021gammaalps1}. This section is organized as follows. 
We first describe the statistical method, then determine the benchmark host-galaxy and Milky-Way magnetic-field models, and finally focus on the IGMF effect on the photon survival probability and on the exclusion limits.

It is worth emphasizing one detail of the host-galaxy implementation. HST observations indicate that GRB 221009A is hosted by a peculiar, nearly edge-on galaxy~\cite{levan2023first}. The burst is located close to the central region, with a projected offset of $0.25''$, corresponding to approximately $0.65\,\mathrm{kpc}$. Its distance to the disk plane is approximately $0.44\,\mathrm{kpc}$, and its projected distance parallel to the disk from the host center is approximately $0.48\,\mathrm{kpc}$~\cite{troitsky2024towards}. The host galaxy has an axial ratio $b/a=0.22\pm0.01$. We therefore place the source at Cartesian coordinates $(x,y,z)=(0.48\,\mathrm{kpc},0\,\mathrm{kpc},0.44\,\mathrm{kpc})$, corresponding to the approximation $y_0=0\,\mathrm{kpc}$. The emitted gamma rays then leave the host galaxy along a host-galactic latitude of $b\simeq 12.7^\circ$.

\subsection{Statistical method}
\label{subsec:method}

We use a log-parabolic function,
\begin{equation}
\frac{dN_{\mathrm{int}}}{dE}(J_0,a,b)
= J_0\left(\frac{E}{1\,\mathrm{TeV}}\right)^{-a-b\log(E/1\,\mathrm{TeV})},
\label{eq:intrinsic-spectrum}
\end{equation}
to describe the intrinsic GRB spectrum. The three parameters $J_0$, $a$, and $b$ are treated as nuisance parameters. The predicted spectrum at Earth is then
\begin{equation}
\frac{dN}{dE}(m_a,g_{a\gamma\gamma},J_0,a,b)
= P_{\rm ALP}(E;\gamma\to\gamma)\,
\frac{dN_{\mathrm{int}}}{dE}(J_0,a,b),
\label{eq:observed-spectrum-model}
\end{equation}
where $P_{\rm ALP}(E;\gamma\to\gamma)$ is the photon survival probability computed from Eq.~\eqref{eq:2.28}. For fixed $(m_a,g_{a\gamma\gamma})$, we minimize
\begin{equation}
\begin{aligned}
\chi^2(m_a,g_{a\gamma\gamma};J_0,a,b)
&=\sum_{i=1}^{N}
\frac{\left[\left.\dfrac{dN}{dE}(m_a,g_{a\gamma\gamma};J_0,a,b)\right|_i
-\left.\dfrac{dN}{dE}\right|_{\rm obs,i}\right]^2}{\sigma_i^2},\\
\chi^2(m_a,g_{a\gamma\gamma})
&=\min_{J_0,a,b}\chi^2(m_a,g_{a\gamma\gamma};J_0,a,b),
\end{aligned}
\label{eq:chi2-definition}
\end{equation}
where $dN/dE|_{\rm obs,i}$ and $\sigma_i$ are the observed flux and uncertainty in the $i$-th energy bin. For each fixed $m_a$, we define
\begin{equation}
\Delta\chi^2(m_a,g_{a\gamma\gamma})
=\chi^2(m_a,g_{a\gamma\gamma})
-\min_{g_{a\gamma\gamma}}\chi^2(m_a,g_{a\gamma\gamma}).
\label{eq:delta-chi2}
\end{equation}
Assuming Wilks' theorem, $\Delta\chi^2=2.71$ gives the one-sided 95\% confidence-level upper limit on $g_{a\gamma\gamma}$ for each value of $m_a$. Throughout this work, we focus on the ALP parameter region $m_a\in[10^{-11},10^{-6}]\,\mathrm{eV}$ and $g_{a\gamma\gamma}\in[10^{-12},10^{-10}]\,\mathrm{GeV}^{-1}$.

\subsection{Host-galaxy and Milky-Way magnetic-field dependence}
\label{subsec:hgmf-mwmf}
We first isolate the effect of the host-galaxy and Milky-Way magnetic-field models by switching off ALP-photon oscillations in intergalactic space and retaining only the conventional EBL attenuation in that region.

\begin{figure}[htbp]
\centering
\includegraphics[width=0.75\textwidth]{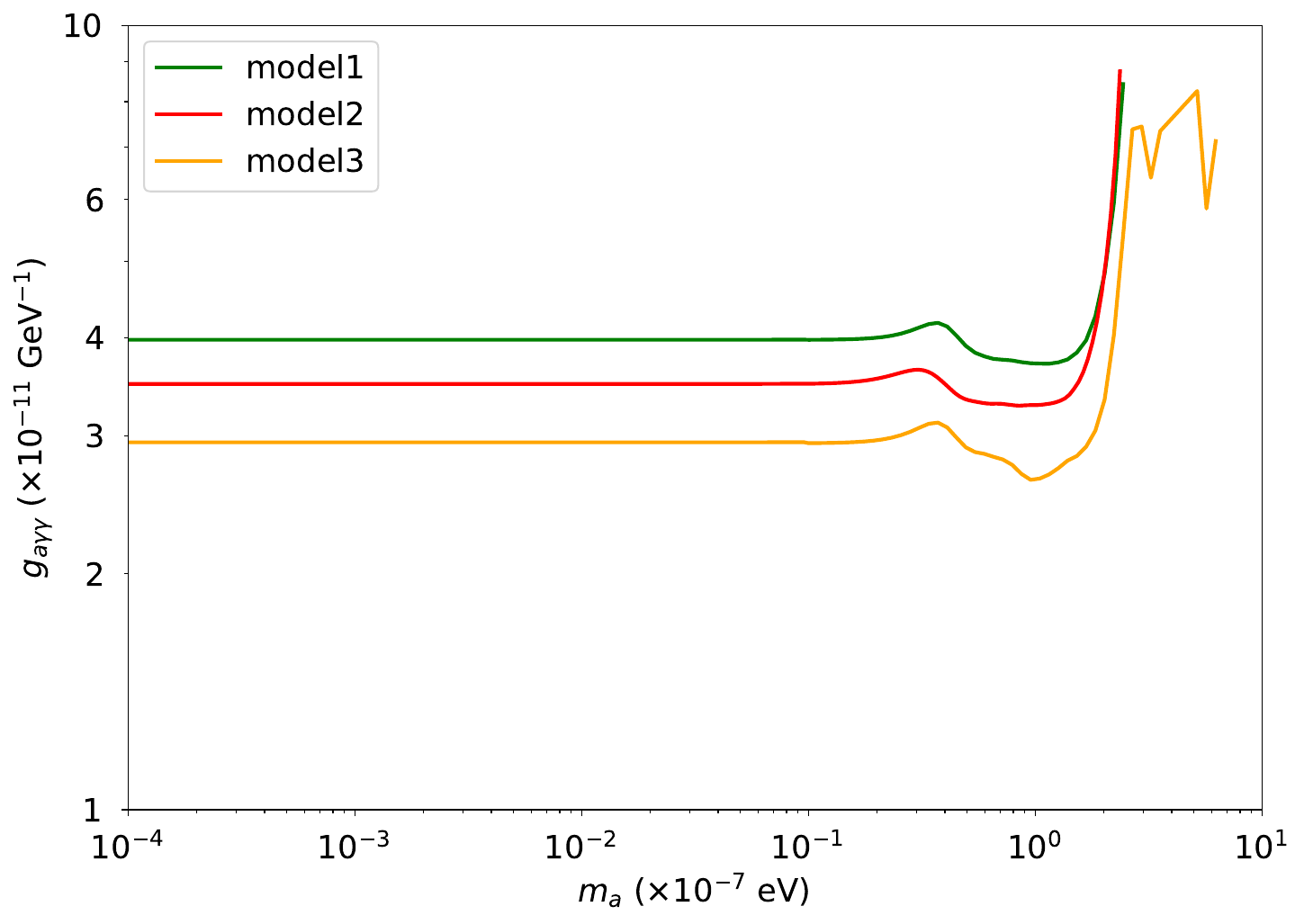}
\qquad
\includegraphics[width=0.75\textwidth]{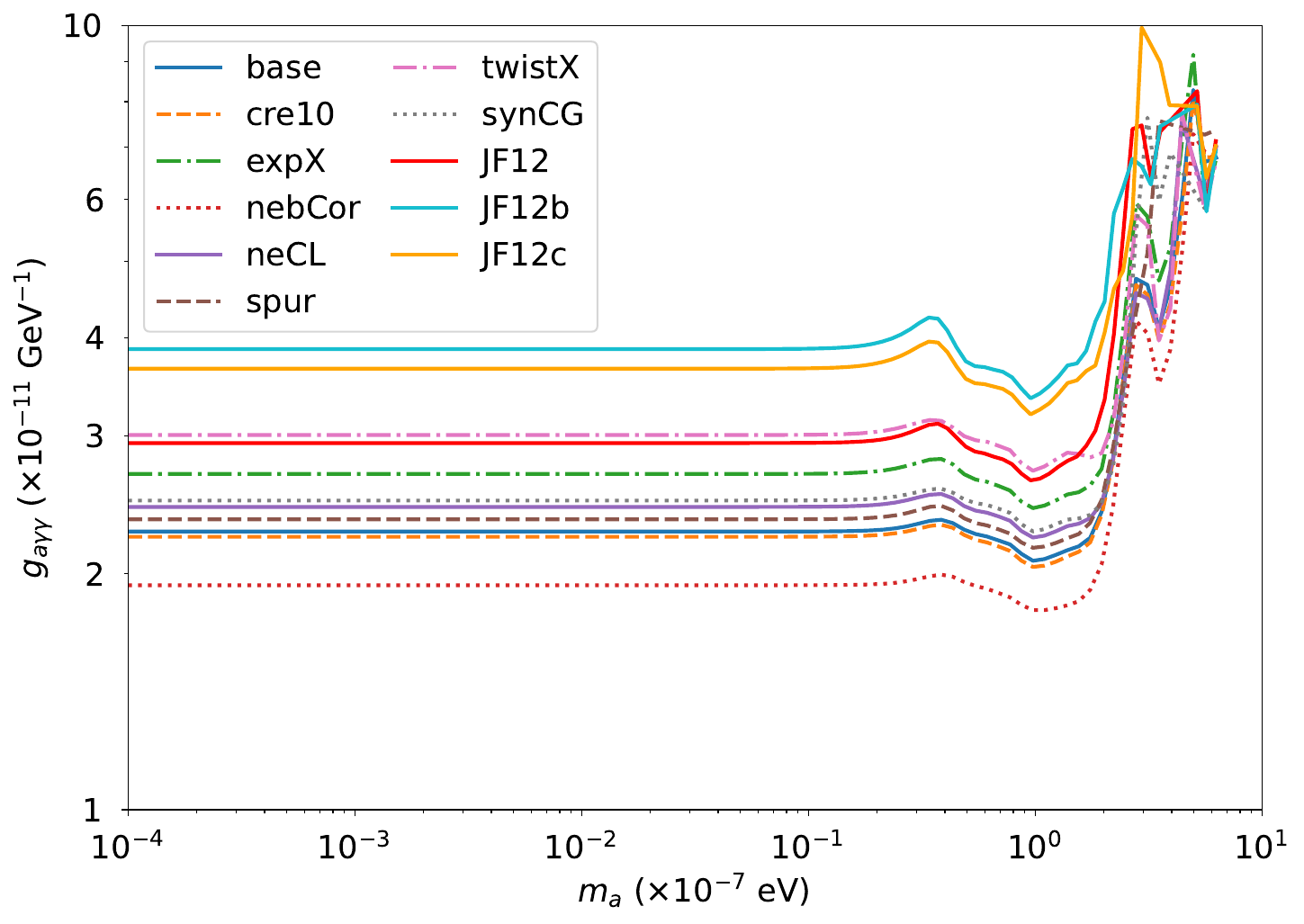}
\caption{\textbf{Top:} The 95\% C.L. exclusion limits from GRB~221009A obtained with different HGMF models, with the MWMF model fixed to JF12. \textbf{Bottom:} The 95\% C.L. exclusion limits obtained with different MWMF models, with the HGMF model fixed to Model~3. In both panels, the region above each curve is excluded.}
\label{fig:A2}
\end{figure}
Figure~\ref{fig:A2} compares the 95\% C.L. exclusion limits obtained with different magnetic-field models. In both panels, each curve represents the corresponding exclusion limit in the $(m_a,g_{a\gamma\gamma})$ plane, with the region above the curve excluded. The top panel shows the dependence on the HGMF model, with the MWMF model fixed to JF12. For $m_a\lesssim3\times10^{-8}\,\mathrm{eV}$, the three HGMF models give approximately mass-independent limits at the level of $g_{a\gamma\gamma}\sim(3.0$--$4.0)\times10^{-11}\,\mathrm{GeV}^{-1}$. Among them, Model~3 gives the strongest constraint, reaching roughly $g_{a\gamma\gamma}\simeq3.0\times10^{-11}\,\mathrm{GeV}^{-1}$ in this low-mass region. Since this model is based most directly on HST observations of the host galaxy and on the geometry of an NGC~891-like disk system, we adopt it as the benchmark HGMF model.

The bottom panel shows the dependence on the Milky-Way magnetic-field model, with the HGMF model fixed to Model~3. We compare JF12, JF12b, JF12c~\cite{jansson2012new,adam2016planck}, and the recent UF models~\cite{unger2024coherent}. In the low-mass region, the limits are also approximately mass independent, ranging from about $2\times10^{-11}\,\mathrm{GeV}^{-1}$ to $4\times10^{-11}\,\mathrm{GeV}^{-1}$. The UF models generally yield tighter constraints than the JF models. We therefore choose the UF base model as the benchmark MWMF model, since it is the baseline realization among the UF models and gives an intermediate constraint within that family.

At larger masses, around $m_a\gtrsim10^{-7}\,\mathrm{eV}$, the limits begin to weaken and develop model-dependent structures in both panels. Overall, when measured relative to the benchmark setup adopted below, namely Model~3 for the HGMF and the UF base model for the MWMF, the variations induced by changing the HGMF or MWMF model are moderate compared with the IGMF-induced effects discussed below. Accordingly, in the rest of the analysis, we fix the HGMF to Model~3 and the MWMF to the UF base model, and focus on the IGMF uncertainty.

\subsection{Impact of the IGMF on photon propagation}
\label{subsec:igmf-survival}
In this subsection, we focus on the effect of ALP--photon oscillations during intergalactic propagation. To isolate this effect, we take the initial state at the entrance of the intergalactic region to be an unpolarized photon beam, $\rho_0=\mathrm{diag}(1/2,1/2,0)$, for both the oscillation and no-oscillation cases, rather than using the output state after host-galaxy propagation. Unless otherwise stated, the IGMF is modeled as a sequence of magnetized domains with transverse field strength $B_T=1\,\mathrm{nG}$, domain length $L_{\rm dom}=1\,\mathrm{Mpc}$ and filling fraction $f=1$, with randomly oriented field directions. We first compare the DLSME model with different smoothing parameters $\sigma$. The limit $\sigma=0$ corresponds to the DLSHE model.

\begin{figure}[htbp]
\centering
\includegraphics[width=0.46\textwidth]{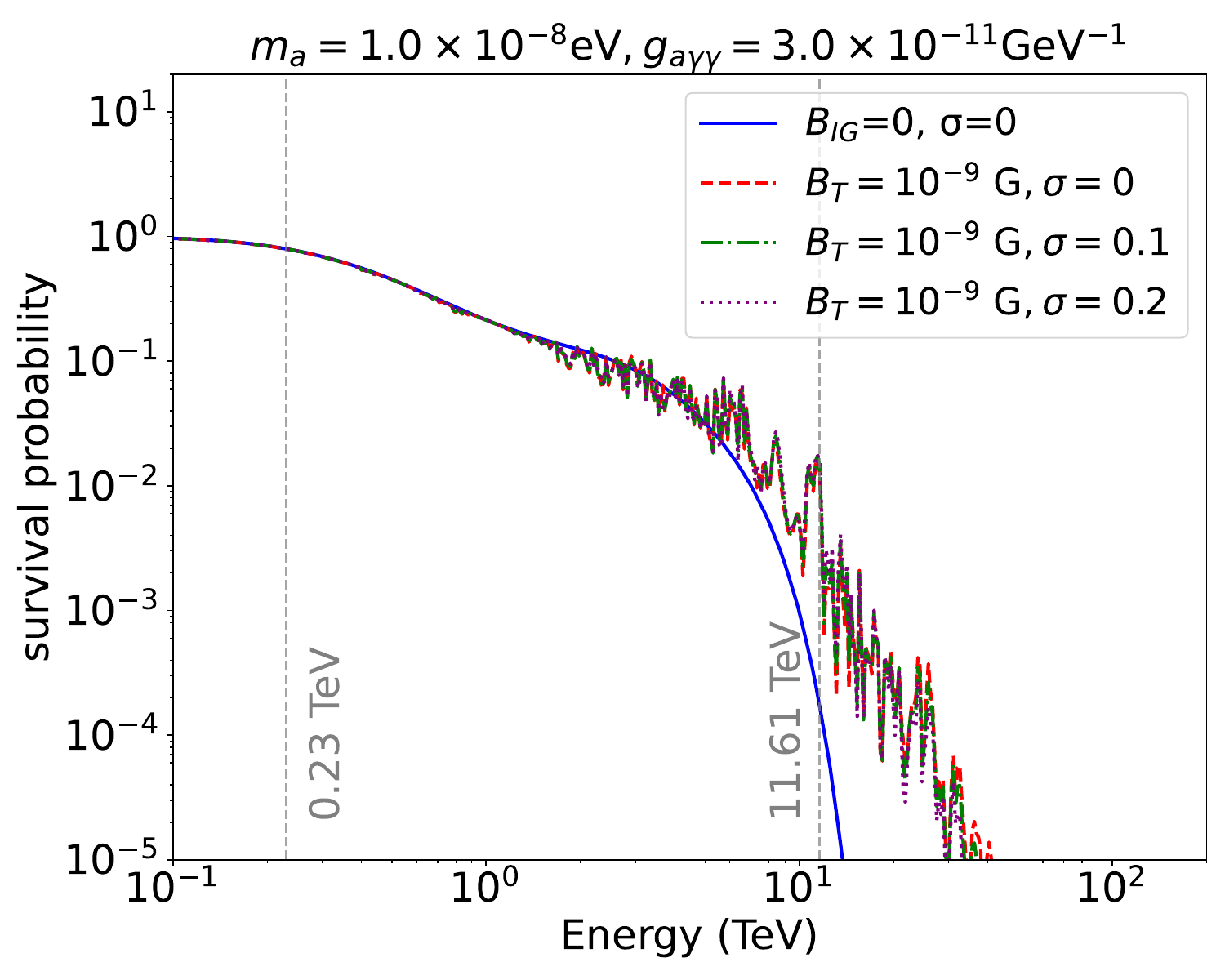}
\qquad
\includegraphics[width=0.46\textwidth]{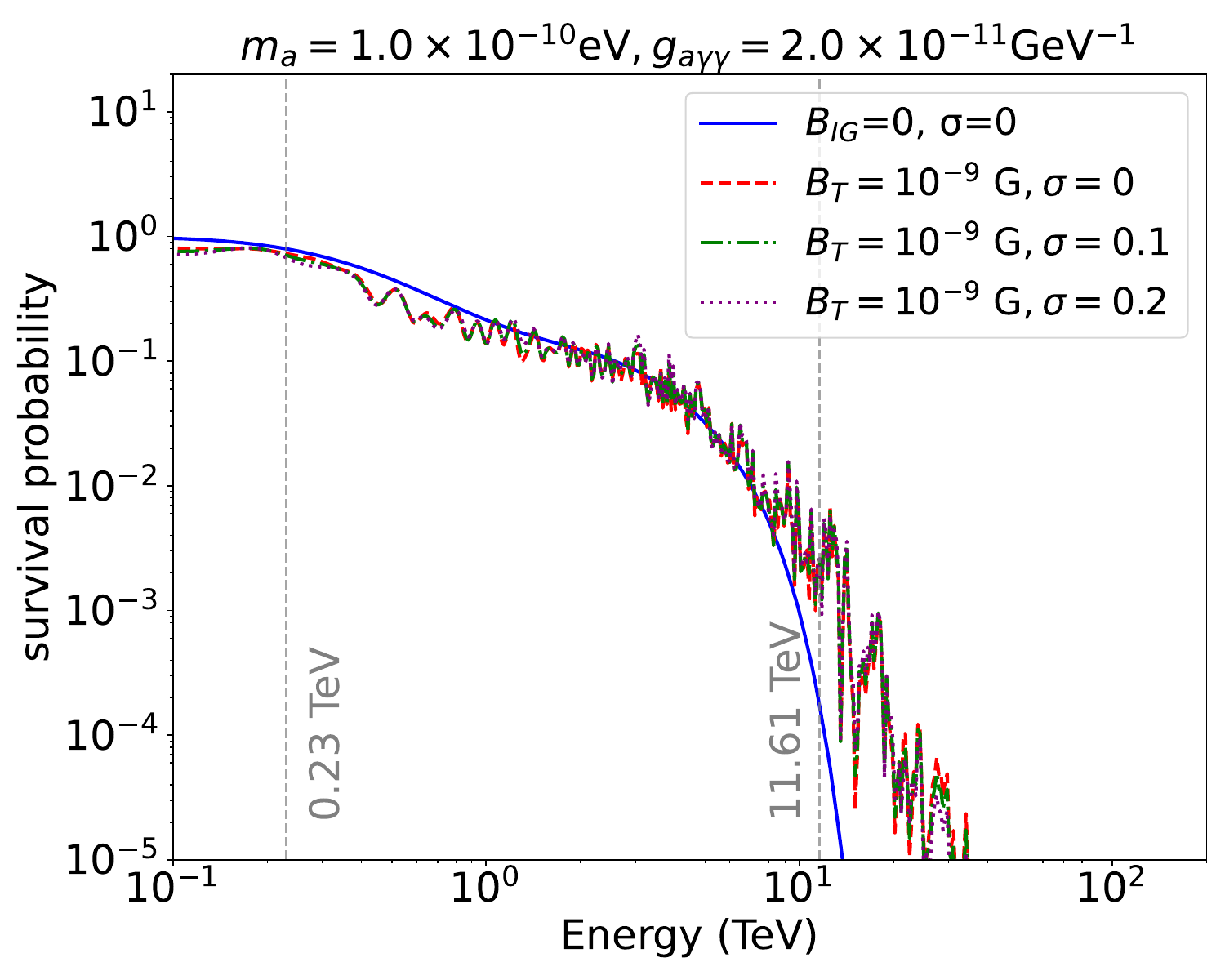}
\caption{Photon survival probability for GRB~221009A, comparing the conventional EBL attenuated case ($B_{\rm IG}=0$) with the results obtained after including the IGMF modeled with the DLSME prescription for different values of the smoothing parameter $\sigma$. The IGMF has $B_T=10^{-9}\,\mathrm{G}$, $L_{\rm dom}=1\,\mathrm{Mpc}$, and filling factor $f=1$. The left panel uses $m_a=1.0\times10^{-8}\,\mathrm{eV}$ and $g_{a\gamma\gamma}=3.0\times10^{-11}\,\mathrm{GeV}^{-1}$, while the right panel uses $m_a=1.0\times10^{-10}\,\mathrm{eV}$ and $g_{a\gamma\gamma}=2.0\times10^{-11}\,\mathrm{GeV}^{-1}$.}
\label{fig:survival-sigma}
\end{figure}

Figure~\ref{fig:survival-sigma} shows the photon survival probability for GRB~221009A with and without ALP--photon oscillations in the IGMF. The blue curve in each panel represents the conventional EBL-attenuated case with $B_{\rm IG}=0$, while the other curves include the IGMF described by the DLSME model with smoothing parameters $\sigma=0$, $0.1$, and $0.2$. The left panel uses $m_a=1.0\times10^{-8}\,\mathrm{eV}$ and $g_{a\gamma\gamma}=3.0\times10^{-11}\,\mathrm{GeV}^{-1}$, whereas the right panel uses $m_a=1.0\times10^{-10}\,\mathrm{eV}$ and $g_{a\gamma\gamma}=2.0\times10^{-11}\,\mathrm{GeV}^{-1}$. The two gray vertical dashed lines mark the energy interval adopted in this analysis, from $0.23\,\mathrm{TeV}$ to $11.61\,\mathrm{TeV}$.

The figure shows that including the IGMF can significantly enhance the photon survival probability at high energies relative to the conventional $B_{\rm IG}=0$ case. This enhancement is most visible near the upper end of the LHAASO energy range, where EBL absorption strongly suppresses the photon flux in the absence of intergalactic ALP--photon oscillations. This occurs because photons can convert into ALPs during propagation through the IGMF; these ALPs are not attenuated by EBL absorption and can later reconvert into photons, thereby increasing the final photon survival probability. 

As suggested by the oscillation-length behavior illustrated in Fig.~\ref{fig:1}, for the ALP benchmark used in the left panel of Fig.~\ref{fig:survival-sigma}, the formal validity condition of the DLSHE approximation, $l_{\rm osc}\gg L_{\rm dom}$, is not uniformly satisfied over the energy interval adopted in this analysis. Nevertheless, the dependence on the DLSME smoothing parameter remains weak. Within each panel, changing $\sigma$ from $0$ to $0.1$ and $0.2$ only produces a minor change in the survival probability.

\subsection{IGMF effects on the exclusion limits}
\label{subsec:igmf-constraints}
\begin{figure}[htbp]
\centering
\includegraphics[width=0.75\textwidth]{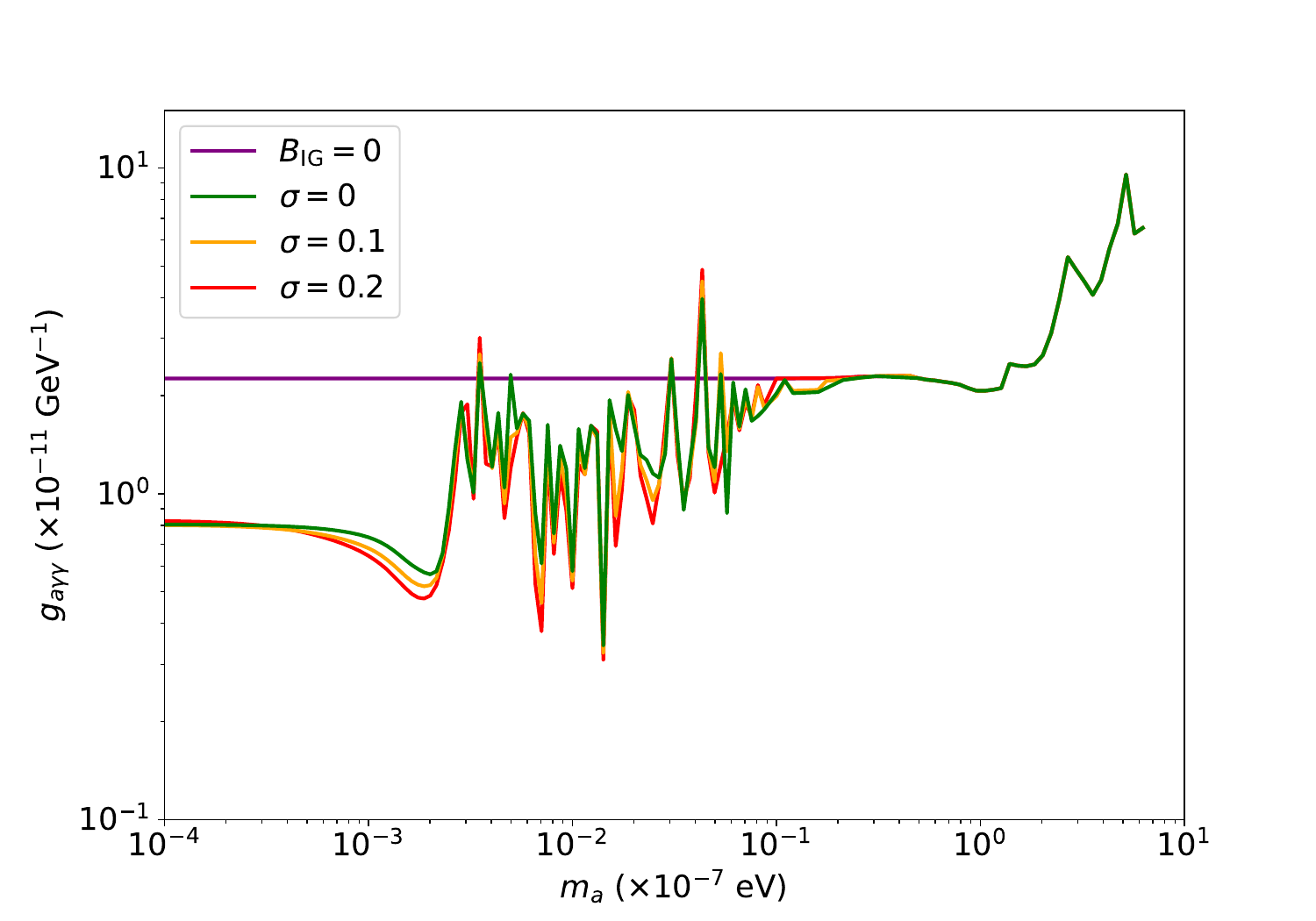}
\caption{The 95\% C.L. exclusion limits from GRB~221009A, obtained without the IGMF ($B_{\rm IG}=0$) and after including the IGMF modeled with the DLSME prescription for different values of the smoothing parameter $\sigma$. The region above each curve is excluded. The IGMF setup uses a domain length $L_{\rm dom}=1\,\mathrm{Mpc}$ and filling factor $f=1$.}
\label{fig:sigma-constraints}
\end{figure}

Figure~\ref{fig:sigma-constraints} shows the ALP constraints obtained without the IGMF ($B_{\rm IG}=0$) and with the DLSME model for $\sigma=0$, $0.1$, and $0.2$, using a domain length $L_{\rm dom}=1\,\mathrm{Mpc}$ and filling factor $f=1$. Each curve represents the corresponding 95\% C.L. exclusion limit in the $(m_a,g_{a\gamma\gamma})$ plane, with the region above the curve excluded. Varying $\sigma$ within $[0,0.2]$ does not substantially change the resulting constraints. This is consistent with the behavior of the photon survival probability discussed in Sec.~\ref{subsec:igmf-survival}, where the dependence on $\sigma$ was found to be subdominant in the considered energy range. We therefore use the DLSHE approximation in the subsequent ALP parameter scan and derive the corresponding exclusion limits for different choices of $L_{\rm dom}$. This approximation significantly reduces the computational cost. Moreover, the simple domain structure of the DLSHE model also facilitates the analytic discussion of the IGMF effect in Sec.~\ref{subsec:physical-origin}.

Figure~\ref{fig:sigma-constraints} also illustrates the main qualitative impact of the IGMF on the exclusion limits. 
In the conventional $B_{\rm IG}=0$ case, the exclusion limit is nearly independent of $m_a$ for $m_a\lesssim3\times10^{-8}\,\mathrm{eV}$, approaching $g_{a\gamma\gamma}\simeq2.3\times10^{-11}\,\mathrm{GeV}^{-1}$. This behavior follows from the fact that the photon survival probability is controlled mainly by photon--ALP conversion in the host galaxy and in the Milky Way. For small $m_a$, the galactic oscillation length is much larger than the galactic coherence length, $l_{\rm osc}\gg\lambda_{\rm coh}$, so the conversion probability becomes approximately independent of $m_a$. A detailed explanation of this mass-independent behavior is given in Sec.~\ref{subsec:physical-origin}.

Once the IGMF is included, the exclusion limits display three qualitatively different mass regions. For relatively large ALP masses, roughly $m_a\gtrsim10^{-8}\,\mathrm{eV}$ for the benchmark choices shown here, photon--ALP conversion in intergalactic space is inefficient, and the limits remain close to the $B_{\rm IG}=0$ result. For intermediate masses, roughly $10^{-10}\,\mathrm{eV}\lesssim m_a\lesssim10^{-8}\,\mathrm{eV}$, the IGMF conversion becomes efficient and the limits develop oscillatory structures. For sufficiently small masses, roughly $m_a\lesssim10^{-10}\,\mathrm{eV}$ in our benchmark examples, the oscillatory dependence on $m_a$ disappears, while the IGMF still enhances the high-energy photon survival probability and therefore gives stronger constraints. These numerical boundaries should be understood as approximate ranges for the parameter choices considered here; in particular, the upper transition depends on $L_{\rm dom}$.
\begin{figure}[htbp]
\centering
\includegraphics[width=0.75\textwidth]{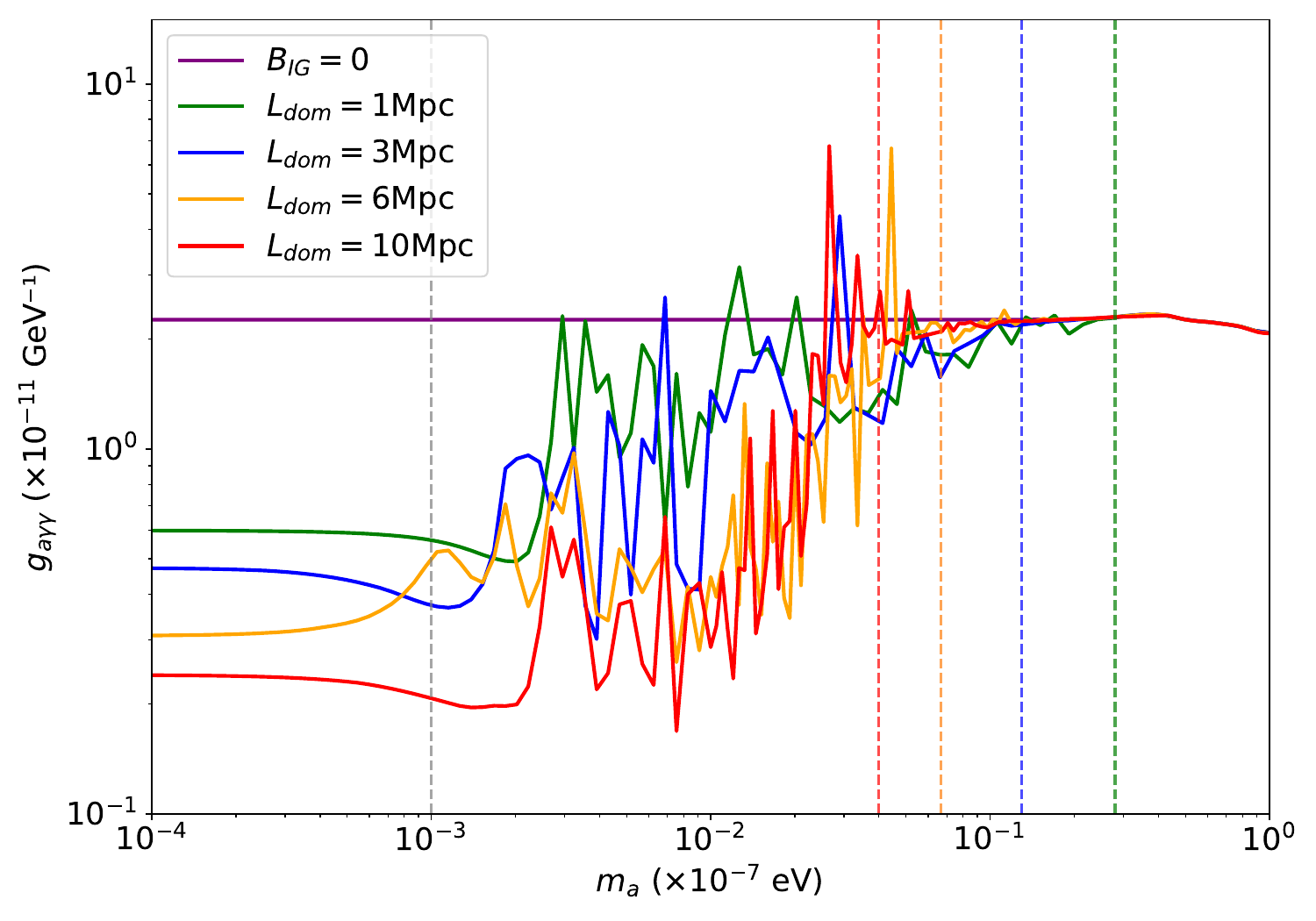}
\caption{The 95\% C.L. exclusion limits from GRB~221009A without the IGMF ($B_{\rm IG}=0$) and with the IGMF included for different magnetic-domain lengths $L_{\rm dom}$, with filling factor fixed to $f=1$. The region above each curve is excluded. The colored dashed lines indicate the onset of significant IGMF-induced oscillatory behavior for each value of $L_{\rm dom}$, while the gray dashed line marks the approximate low-mass end of the oscillatory region.}
\label{fig:Ldom-constraints}
\end{figure}

Figure~\ref{fig:Ldom-constraints} shows the dependence of the 95\% C.L. exclusion limits on the IGMF domain length $L_{\rm dom}$, for representative values $1$, $3$, $6$, and $10\,\mathrm{Mpc}$. Each solid curve represents the corresponding exclusion limit in the $(m_a,g_{a\gamma\gamma})$ plane, with the region above the curve excluded. The results are obtained using the DLSHE approximation. The colored dashed lines indicate, for each value of $L_{\rm dom}$, the approximate mass at which significant IGMF-induced oscillations begin to appear. The gray dashed line marks the approximate low-mass end of the oscillatory behavior. Changing $L_{\rm dom}$ mainly shifts the onset of the oscillatory region, while the low-mass boundary is comparatively less affected in the benchmark cases shown here. A more detailed physical interpretation is given in Sec.~\ref{subsec:physical-origin}.

\subsection{Physical origin of the exclusion limits}
\label{subsec:physical-origin}

We now explain the physical origin of the three mass regions identified above. The essential control parameter is the ratio between the oscillation length $l_{\rm osc}$ and the magnetic-domain length $L_{\rm dom}$. This ratio determines both the efficiency of photon--ALP conversion in each domain and the phase accumulated during multi-domain propagation.

In a single homogeneous magnetic domain, the photon--ALP conversion amplitude is governed by the off-diagonal part of the transfer matrix,
\begin{equation}
\mathcal{A}_{\gamma a}
\simeq
\frac{l_{\rm osc}\Delta_{a\gamma}}{\pi}
\sin\left(\frac{\pi L_{\rm dom}}{l_{\rm osc}}\right).
\label{eq:offdiag-physical}
\end{equation}
The sine factor describes the oscillatory phase accumulated across one domain, whereas the prefactor $l_{\rm osc}\Delta_{a\gamma}/\pi$ controls the amplitude of the mixing. This distinction is important: rapid phase variation alone does not imply efficient conversion if the prefactor is small.

To see this explicitly, we consider, for simplicity, an initial density matrix at the entrance of the IGMF given by $\rho_0=\mathrm{diag}(1/2,1/2,0)$ and neglect EBL absorption, so that the following discussion isolates the intrinsic multi-domain photon--ALP oscillation effect. This simplified setup is sufficient to explain the physical origin of the three mass regions discussed above. We then consider the density matrix after crossing the $n$-th IGMF domain,
\begin{equation}
\rho^{(n)}=\mathcal{U}_n\rho^{(n-1)}\mathcal{U}_n^\dagger .
\label{eq:rho-domain-evolution}
\end{equation}
Within one domain, the mixing matrix is approximately constant, so that
\begin{equation}
\mathcal{U}_n=\exp\left(i\mathcal{M}L_{\rm dom}\right).
\label{eq:domain-transfer}
\end{equation}
The mixing matrix can be written as
\begin{equation}
\mathcal{M}=V^\dagger(\phi_n)\mathcal{M}_0^{(n)}V(\phi_n),
\label{eq:rotated-mixing-matrix}
\end{equation}
where $\phi_n$ is the angle between the transverse magnetic field $\mathbf{B}_T$ and the $z$-axis, and
\begin{equation}
V(\phi_n)=
\begin{pmatrix}
\cos\phi_n & -\sin\phi_n & 0\\
\sin\phi_n & \cos\phi_n & 0\\
0 & 0 & 1
\end{pmatrix}.
\label{eq:rotation-matrix-domain}
\end{equation}
When discussing the intrinsic oscillation in one domain, we neglect absorption and use the fact that QED birefringence is negligible for the weak IGMF considered here. $\Delta_\perp$ and $\Delta_\parallel$ can then be approximated by their average,
\begin{equation}
\bar{\Delta}=\frac{\Delta_\perp+\Delta_\parallel}{2},
\end{equation}
and the mixing matrix becomes
\begin{equation}
\mathcal{M}_0^{(n)}\simeq
\begin{pmatrix}
\bar{\Delta} & 0 & 0\\
0 & \bar{\Delta} & \Delta_{a\gamma}\\
0 & \Delta_{a\gamma} & \Delta_{aa}
\end{pmatrix}.
\label{eq:approx-domain-mixing}
\end{equation}
The nontrivial part is the $2\times2$ photon--ALP block,
\begin{equation}
\mathcal{U}_{2\times2}^{(n)}
=
\exp\left(i\mathcal{M}_{2\times2}^{(n)}L_{\rm dom}\right),
\qquad
\mathcal{M}_{2\times2}^{(n)}=
\begin{pmatrix}
\bar{\Delta} & \Delta_{a\gamma}\\
\Delta_{a\gamma} & \Delta_{aa}
\end{pmatrix}.
\label{eq:two-by-two-block}
\end{equation}
Defining
\begin{equation}
\Delta=\bar{\Delta}-\Delta_{aa},
\qquad
\Delta_{\rm osc}=\sqrt{\Delta^2+4\Delta_{a\gamma}^2},
\qquad
l_{\rm osc}=\frac{2\pi}{\Delta_{\rm osc}},
\label{eq:delta-osc-def}
\end{equation}
one obtains
\begin{equation}
\begin{aligned}
\mathcal{U}_{2\times2}^{(n)}
&=
e^{i(\bar{\Delta}+\Delta_{aa})L_{\rm dom}/2}
\left[
\cos\left(\frac{\pi L_{\rm dom}}{l_{\rm osc}}\right)\mathbf{1}
\right.
\\
&\hspace{2.0cm}
\left.
+i\sin\left(\frac{\pi L_{\rm dom}}{l_{\rm osc}}\right)
\begin{pmatrix}
l_{\rm osc}\Delta/2\pi & l_{\rm osc}\Delta_{a\gamma}/\pi\\
l_{\rm osc}\Delta_{a\gamma}/\pi & -l_{\rm osc}\Delta/2\pi
\end{pmatrix}
\right].
\end{aligned}
\label{eq:two-by-two-transfer}
\end{equation}
The off-diagonal element of Eq.~\eqref{eq:two-by-two-transfer} gives the photon--ALP conversion amplitude $\mathcal{A}_{\gamma a}$ in Eq.~\eqref{eq:offdiag-physical}. Since Eq.~\eqref{eq:rho-domain-evolution} evolves the density matrix from $\rho^{(n-1)}$ to $\rho^{(n)}$ through the transfer matrix $\mathcal{U}_n$, the change in the ALP component $\rho_{33}^{(n)}$ across one domain is mainly governed by the size of $\mathcal{A}_{\gamma a}$. After propagation through many IGMF domains, the final ALP component is determined by the ordered product of the corresponding transfer matrices. Thus, the IGMF contribution to the photon survival probability is controlled by both the single-domain conversion amplitude and its accumulation over many domains.

\begin{figure}[htbp]
\centering
\includegraphics[width=0.75\textwidth]{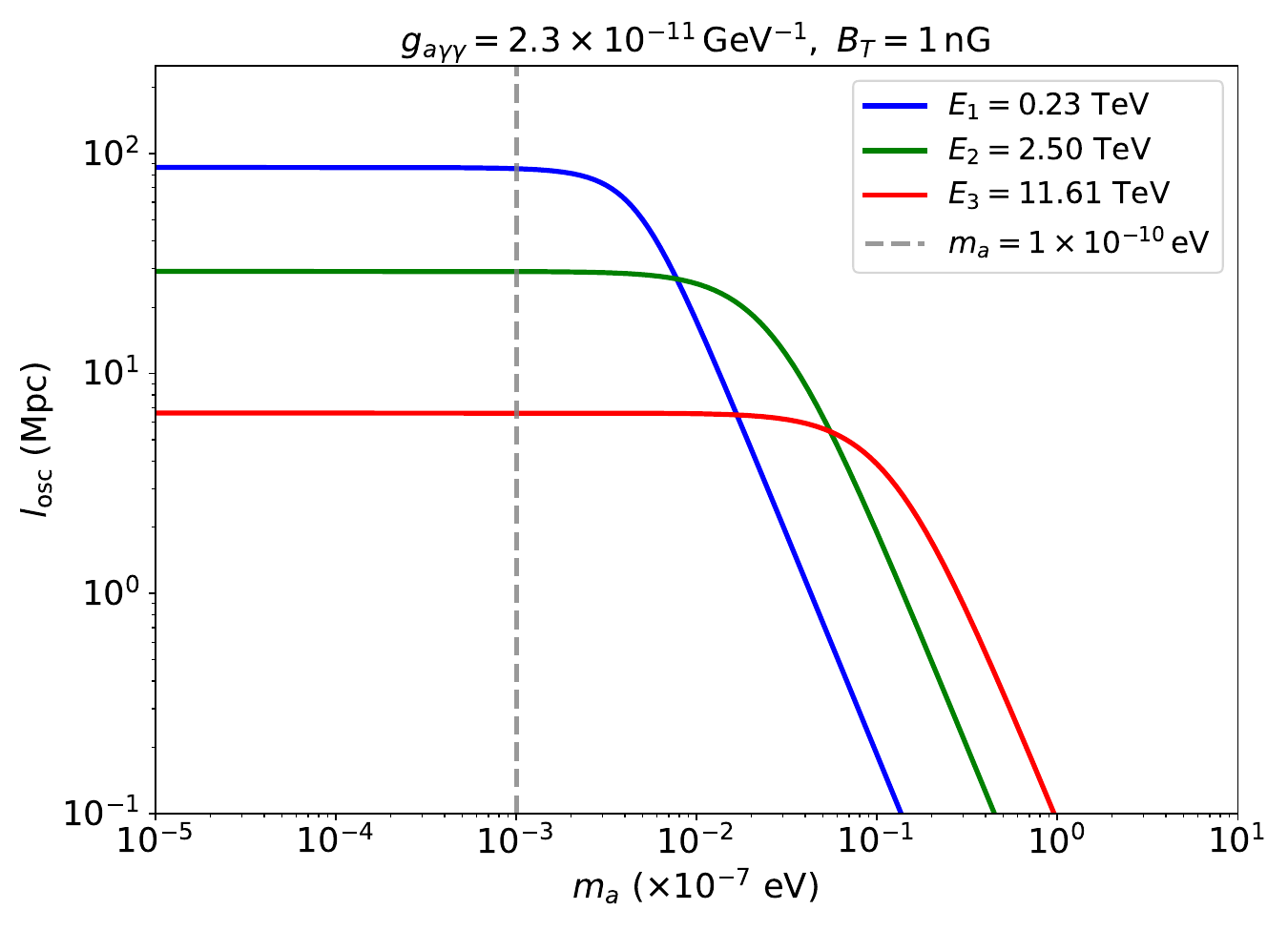}
\caption{Mass dependence of the oscillation length $l_{\rm osc}(E)$ for three representative photon energies, $E=0.23$, $2.50$, and $11.61\,{\rm TeV}$. The benchmark parameters are $B_T=1\,{\rm nG}$, $g_{a\gamma\gamma}=2.3\times10^{-11}\,{\rm GeV}^{-1}$, and $n_e\sim10^{-7}\,{\rm cm}^{-3}$. The gray dashed line at $m_a=1\times10^{-10}\,{\rm eV}$ marks the approximate low-mass transition below which $l_{\rm osc}$ becomes nearly independent of $m_a$.}
\label{fig:single-domain-scales}
\end{figure}

Since $\mathcal{A}_{\gamma a}$ depends explicitly on $l_{\rm osc}$ through Eq.~\eqref{eq:offdiag-physical}, its mass dependence is mainly inherited from the mass dependence of $l_{\rm osc}$. It is therefore useful to examine how $l_{\rm osc}$ varies with $m_a$ in the IGMF setup. Figure~\ref{fig:single-domain-scales} shows this dependence for three representative photon energies, $E=0.23$, $2.50$, and $11.61\,\mathrm{TeV}$, with $B_T=1\,\mathrm{nG}$, $g_{a\gamma\gamma}=2.3\times10^{-11}\,\mathrm{GeV}^{-1}$, and $n_e\sim10^{-7}\,\mathrm{cm}^{-3}$. The gray dashed line marks the approximate low-mass transition below which $l_{\rm osc}$ is nearly independent of $m_a$. This behavior provides the basis for the following classification into the large-mass, intermediate-mass, and small-mass regions.

\paragraph{Large-mass region.}
For relatively large ALP masses, roughly $m_a \gtrsim 10^{-8}\,\mathrm{eV}$ in the benchmark cases shown in Fig.~\ref{fig:Ldom-constraints}, the ALP mass term $\Delta_{aa}$ leads to a short oscillation length, with $l_{\rm osc}\lesssim L_{\rm dom}$. Although the phase $\pi L_{\rm dom}/l_{\rm osc}$ can vary rapidly, the prefactor $l_{\rm osc}\Delta_{a\gamma}/\pi$ is small. 
For example, for $g_{a\gamma\gamma}=2.3\times10^{-11}\,\mathrm{GeV}^{-1}$, the off-diagonal factor is approximately bounded by
\begin{equation}
\left|
\frac{l_{\rm osc}\Delta_{a\gamma}}{\pi}
\sin\left(\frac{\pi L_{\rm dom}}{l_{\rm osc}}\right)
\right|
\lesssim
5\times10^{-3}
\end{equation}
over the energy interval $0.23\,\mathrm{TeV}\leq E\leq11.61\,\mathrm{TeV}$. The corresponding conversion probability per domain is therefore no larger than a few times $10^{-6}$,
\begin{equation}
P_{\gamma\to a}^{\rm domain}
\sim
|\mathcal{A}_{\gamma a}|^2
\lesssim
3\times10^{-6}.
\end{equation}

This is much smaller than the typical photon--ALP conversion probability in the host galaxy or in the Milky Way, which can be of order $10^{-3}$ or larger. Consequently, IGMF-induced conversion is too weak to appreciably modify the photon survival probability. This explains why, in the large-mass region, the exclusion limits with IGMF approach the $B_{\rm IG}=0$ result.

\paragraph{Intermediate-mass region.}
As $m_a$ decreases, $l_{\rm osc}$ increases. In the approximate range $10^{-10}\,\mathrm{eV}\lesssim m_a\lesssim10^{-8}\,\mathrm{eV}$, depending on $L_{\rm dom}$, there can be an energy $E_0$ in the energy interval $0.23\,{\rm TeV}\leq E \leq 11.61\,{\rm TeV}$ such that
\begin{equation}
l_{\rm osc}(E_0)\gtrsim L_{\rm dom}.
\label{eq:losc-Ldom-condition}
\end{equation}
In this regime, the factor $l_{\rm osc}\Delta_{a\gamma}/\pi$ is no longer strongly suppressed, and the off-diagonal element of the transfer matrix can reach the level of $10^{-2}$ or larger. The conversion probability accumulated through many IGMF domains can therefore become sizable. As a result, the IGMF contribution can become comparable to, or larger than, the galactic contribution.

The oscillatory structure of the exclusion limits is a multi-domain effect. After propagation through $n$ randomly oriented domains, the ALP component $\rho_{33}^{(n)}$ contains products of the form
\begin{equation}
\sin^p\omega\cos^q\omega,
\qquad
\omega=\frac{\pi L_{\rm dom}}{l_{\rm osc}},
\qquad
p+q=2n.
\label{eq:sin-cos-products-main}
\end{equation}
Because $l_{\rm osc}$ depends on $m_a$, these products generate a rapidly varying dependence of $\rho_{33}^{(n)}$ on the ALP mass. This is the physical origin of the oscillatory structures in Figs.~\ref{fig:sigma-constraints} and~\ref{fig:Ldom-constraints}. 
The onset of pronounced oscillatory structures can be estimated by the condition $l_{\rm osc}^{\rm max}\simeq L_{\rm dom}$, where $l_{\rm osc}^{\rm max}$ denotes the maximum value of $l_{\rm osc}(E)$ over the energy interval $0.23\,{\rm TeV}\leq E \leq 11.61\,{\rm TeV}$. The colored dashed lines in Fig.~\ref{fig:Ldom-constraints} mark the values of $m_a$ obtained from this condition, corresponding to the onset of pronounced oscillatory behavior for each IGMF domain length.

\paragraph{Small-mass region.}
For sufficiently small ALP masses, roughly $m_a\lesssim10^{-10}\,\mathrm{eV}$ in the benchmark examples, the oscillation length becomes approximately independent of $m_a$, as shown in Fig.~\ref{fig:single-domain-scales}. Consequently, $\pi L_{\rm dom}/l_{\rm osc}$ in the factors $\sin(\pi L_{\rm dom}/l_{\rm osc})$ also becomes nearly independent of $m_a$. The products in Eq.~\eqref{eq:sin-cos-products-main} therefore no longer generate a rapidly varying dependence on $m_a$, which explains why the exclusion limits cease to oscillate in the low-mass region.

The disappearance of oscillations does not mean that the IGMF effect becomes small. $|\mathcal{A}_{\gamma a}|$ remains sizable, and photon--ALP conversion in the IGMF can still produce a substantial ALP population. These ALPs are not attenuated by EBL absorption and can later reconvert into photons, thereby enhancing the high-energy photon survival probability. Therefore, the low-mass exclusion limits become stronger than the $B_{\rm IG}=0$ limits, but their dependence on $m_a$ becomes comparatively smooth. The corresponding transition is indicated by the gray dashed line in Fig.~\ref{fig:Ldom-constraints}, which marks the approximate low-mass end of the oscillatory region. In our numerical examples, this transition occurs at approximately $m_a\simeq10^{-10}\,\mathrm{eV}$.

The mass-independent behavior in the small-mass region has another manifestation. Even when one focuses on the single-domain conversion amplitude in Eq.~\eqref{eq:offdiag-physical}, the limit $l_{\rm osc}\gg L_{\rm dom}$ leads to a cancellation of the explicit $l_{\rm osc}$ dependence. In this limit,
\begin{equation}
\sin\left(\frac{\pi L_{\rm dom}}{l_{\rm osc}}\right)
\simeq
\frac{\pi L_{\rm dom}}{l_{\rm osc}},
\end{equation}
and therefore
\begin{equation}
\mathcal{A}_{\gamma a}
\simeq
\Delta_{a\gamma}L_{\rm dom}.
\end{equation}
Thus, the single-domain conversion probability becomes approximately independent of $l_{\rm osc}$, and hence of $m_a$. This accounts for the low-mass plateau of $P_{\gamma\to a}$ in Fig.~\ref{fig:2}. A similar small-phase mechanism also explains the nearly mass-independent exclusion limit in the $B_{\rm IG}=0$ case for $m_a\lesssim3\times10^{-8}\,\mathrm{eV}$, where the relevant condition is $l_{\rm osc}\gg\lambda_{\rm coh}$ in the host-galaxy and Milky-Way magnetic fields rather than $l_{\rm osc}\gg L_{\rm dom}$ in the IGMF.

\begin{figure}[htbp]
\centering
\includegraphics[width=0.75\textwidth]{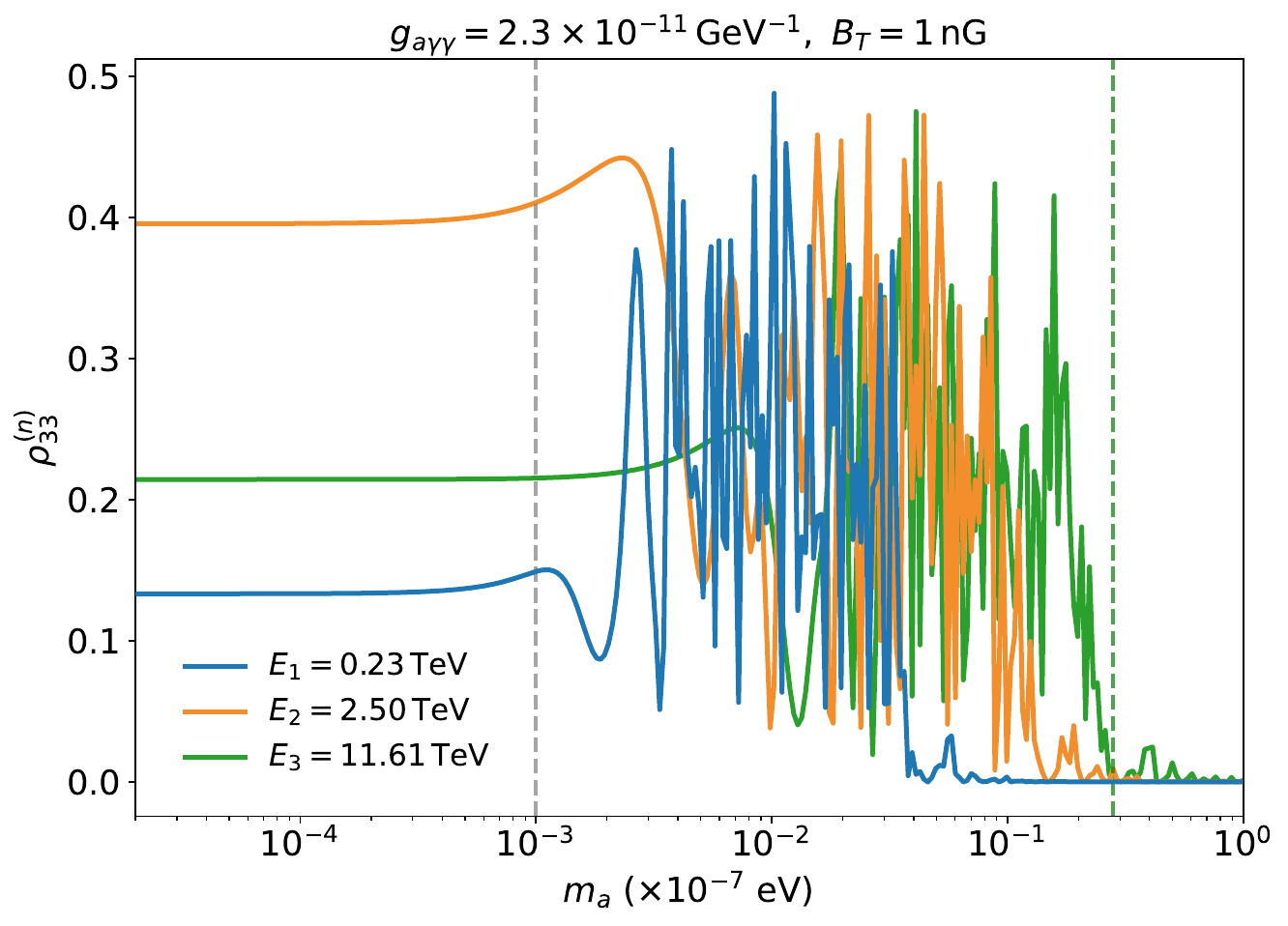}
\caption{Dependence of the ALP component $\rho_{33}^{(n)}$ on the ALP mass $m_a$ after propagation through the IGMF, for different photon energies $E$. The benchmark parameters are $L_{\rm dom}=1\,\mathrm{Mpc}$, $B_T=1\,\mathrm{nG}$, $g_{a\gamma\gamma}=2.3\times10^{-11}\,\mathrm{GeV}^{-1}$, $n_e\sim10^{-7}\,\mathrm{cm}^{-3}$, and $\rho_0=\mathrm{diag}(1/2,1/2,0)$. The right green vertical dashed line marks the approximate transition from the large-mass region to the intermediate-mass region, while the left gray vertical dashed line marks the approximate transition from the intermediate-mass region to the small-mass region.}
\label{fig:rho33-mass}
\end{figure}

The three regimes described above are displayed directly in Fig.~\ref{fig:rho33-mass}. Figure~\ref{fig:rho33-mass} shows the dependence of the accumulated ALP component $\rho^{(n)}_{33}$ on the ALP mass $m_a$ after propagation through the IGMF, for three representative photon energies, $E=0.23$, $2.50$, and $11.61\,\mathrm{TeV}$. We take $L_{\rm dom}=1\,\mathrm{Mpc}$, $B_T=1\,\mathrm{nG}$, $g_{a\gamma\gamma}=2.3\times10^{-11}\,\mathrm{GeV}^{-1}$, $n_e\sim10^{-7}\,\mathrm{cm}^{-3}$, and the initial density matrix $\rho_0=\mathrm{diag}(1/2,1/2,0)$. The right green vertical dashed line marks the approximate transition from the large-mass region to the intermediate-mass region, while the left gray vertical dashed line marks the approximate transition from the intermediate-mass region to the small-mass region. These two lines therefore divide the plotted mass range into the large-mass, intermediate-mass, and small-mass regions discussed above. Unlike the single-domain probability shown in Fig.~\ref{fig:2}, Fig.~\ref{fig:rho33-mass} displays the ALP component accumulated after propagation through many randomly oriented magnetic domains, and therefore illustrates the multi-domain effect directly.

In the large-mass region, the $\gamma\to a$ conversion probability in each IGMF domain is strongly suppressed, so the accumulated ALP component $\rho^{(n)}_{33}$ remains small. In the intermediate-mass region, $\rho^{(n)}_{33}$ varies rapidly with $m_a$, reflecting the multi-domain phase dependence in Eq.~\eqref{eq:sin-cos-products-main}. This rapid mass dependence is inherited by the final photon survival probability after EBL attenuation and gives rise to the oscillatory exclusion contours. 
In the small-mass region, $l_{\rm osc}$ becomes nearly independent of $m_a$, and $|\mathcal{A}_{\gamma a}|$ remains sizable and approximately mass independent. Photon--ALP conversion in the IGMF can therefore build up a substantial ALP component, leading to a sizable $\rho^{(n)}_{33}$ after multi-domain propagation.

The transition from the large-mass regime can also be understood from the single-domain conversion probability in Eq.~\eqref{eq:2.24}. As shown in Fig.~\ref{fig:2}, at large $m_a$ the oscillation length is short, and the prefactor $(g_{a\gamma\gamma}B_T l_{\rm osc}/2\pi)^2$ strongly suppresses the conversion probability. As $m_a$ decreases, $l_{\rm osc}$ increases and gradually becomes comparable to $L_{\rm dom}$. For still smaller $m_a$, the envelope of the conversion probability rises by several orders of magnitude, and the probability increases from the zero at $l_{\rm osc}=L_{\rm dom}$ to sizable conversion values. Therefore, the condition $l_{\rm osc}\sim L_{\rm dom}$ provides a useful estimate of the mass scale beyond which IGMF-induced conversion becomes efficient enough to affect the exclusion limit.

The physical picture in this section also explains the energy dependence shown in Fig.~\ref{fig:survival-sigma}. In the low-energy range, EBL absorption is weak and the total particle number is approximately conserved, so a larger final ALP population corresponds to a smaller final photon population. In the high-energy range, EBL absorption strongly suppresses photons, and the surviving photons are mainly supplied by ALPs that were produced earlier and later reconvert into photons. In this regime, the photon survival probability is positively correlated with the ALP population generated during propagation.

In summary, the shape of the exclusion limits is determined by the interplay between $l_{\rm osc}$, $L_{\rm dom}$, and EBL absorption. At relatively large masses, roughly $m_a\gtrsim10^{-8}\,\mathrm{eV}$ for our benchmark choices, the IGMF conversion amplitude is suppressed, so the limits reduce to the $B_{\rm IG}=0$ result. At intermediate masses, roughly $10^{-10}\,\mathrm{eV}\lesssim m_a\lesssim10^{-8}\,\mathrm{eV}$, multi-domain conversion becomes efficient and the dependence of $\rho_{33}^{(n)}$ on $m_a$ generates oscillatory exclusion limits. At sufficiently small masses, roughly $m_a\lesssim10^{-10}\,\mathrm{eV}$, $l_{\rm osc}$ becomes nearly mass independent, so the oscillations disappear, but the IGMF still enhances the high-energy photon survival probability and leads to stronger constraints.

\section{Conclusion}
\label{sec:conclude}

We have studied the impact of line-of-sight magnetic fields on ALP constraints derived from the LHAASO observation of GRB 221009A. The propagation setup includes the host-galaxy magnetic field, the intergalactic magnetic field, and the Milky-Way magnetic field. By comparing different magnetic-field models, we find that the host-galaxy and Galactic magnetic-field uncertainties produce only moderate changes in the exclusion limits. The dominant model dependence instead arises from the IGMF.

When the IGMF is included, photon--ALP oscillations in intergalactic space can enhance the high-energy photon survival probability. The resulting exclusion limits differ substantially from those obtained with EBL attenuation alone. In particular, the limits display three regimes. At relatively large ALP mass, roughly $m_a\gtrsim10^{-8}\,\mathrm{eV}$ for the benchmark choices considered here, IGMF-induced conversion is suppressed and the limits approach the $B_{\rm IG}=0$ result. At intermediate mass, roughly $10^{-10}\,\mathrm{eV}\lesssim m_a\lesssim10^{-8}\,\mathrm{eV}$, multi-domain propagation produces oscillatory features in the exclusion contours. At sufficiently small mass, roughly $m_a\lesssim10^{-10}\,\mathrm{eV}$, the oscillation length becomes nearly mass independent, the oscillations disappear, and the IGMF-enhanced survival probability leads to stronger constraints.

The physical origin of these regimes is controlled by the ratio between the oscillation length and the magnetic-domain size. For large $m_a$, the prefactor $l_{\rm osc}\Delta_{a\gamma}/\pi$ in the off-diagonal transfer-matrix element is small, so the IGMF conversion probability per domain is strongly suppressed. Once $l_{\rm osc}$ becomes comparable to $L_{\rm dom}$ within the energy interval adopted in this analysis, the multi-domain transfer matrix generates terms involving products of $\sin\omega$ and $\cos\omega$, with $\omega=\pi L_{\rm dom}/l_{\rm osc}$. This produces the oscillatory mass dependence of the ALP component and therefore of the exclusion limits. In the low-mass region, $l_{\rm osc}$ becomes nearly independent of $m_a$, so the oscillatory behavior ceases, while the conversion amplitude remains sizable.

These results show that realistic modeling of the IGMF is essential for robust ALP constraints from GRB 221009A and, more generally, from long-distance very-high-energy gamma-ray propagation. Future improvements in IGMF modeling, including the field strength, coherence length, filling factor, and directional stochasticity, will be important for reducing the dominant astrophysical uncertainty in such constraints.

\vspace*{3mm}
\noindent 
\textbf{Acknowledgments}\\[0.5mm] 
We acknowledge Zhaohuan Yu for helpful discussions. This work is supported by the National Key R{\&}D Program of China under grant 2023YFA1606100 and by the National Natural Science Foundation of China under grants No. 12435005. C.\,H.\ acknowledges supports from the Sun Yat-Sen University Science Foundation, 
the Fundamental Research Funds for the Central Universities at Sun Yat-sen University under Grant No.\,24qnpy117, and the Key Laboratory of Particle Astrophysics and Cosmology (MOE) of Shanghai Jiao Tong University.

\bibliographystyle{JHEP}
\bibliography{biblio}
\end{document}